\documentclass[12pt,preprint]{aastex}
\usepackage{amsmath}
\usepackage{graphicx}
\usepackage{amsmath,bm}

\begin{document}
\begin{tiny}

\end{tiny}
\title{On the fallback disk around the slowest isolated pulsar, 1E 161348$-$5055}
%\title{The fallback disk mass of the pulsar in RCW103}
\author{Kun Xu$^{1}$ and Xiang-Dong Li$^{1,2}$}
\affil{$^{1}$Department of Astronomy, Nanjing University, Nanjing 210023, China;
lixd@nju.edu.cn}

\affil{$^{2}$Key Laboratory of Modern Astronomy and Astrophysics, Nanjing University,
Ministry of Education, Nanjing 210023, China}

\begin{abstract}

The central compact object 1E 161348$-$5055 in the supernova remnant RCW 103 has a spin period $\sim 6.67$ hr, making it the slowest isolated pulsar. It is believed that a supernova fallback disk is required to spin down the neutron star to the current spin period within a few $10^3$ yr. The mass of the fallback disk around newborn neutron stars can provide useful information on the supernova processes and the possible detection limit with optical/infrared observations. However, it is controversial how massive the disk is in the case of 1E 161348$-$5055. In this work we simulate the spin evolution of a magnetar that is driven by the interaction between the disk and the star's magnetic field. Compared with previous studies, we take into account various critical conditions that affect the formation and evolution of the fallback disk.
Our calculation shows that we can reproduce the extremely slow spin of 1E 161348$-$5055 when taking the initial disk mass $M_{\rm d} \sim 10^{-7} M_{\odot}$ and the neutron star magnetic field $B\geq 5 \times 10^{15}$ G.
This implies that 1E 161348$-$5055 may be a magnetar with very special initial parameters. However, if future observations reveal more objects like 1E 161348$-$5055, then stringent constraints can be obtained on the supernova fallback.

\end{abstract}


\section{Introduction}
The central compact object (CCO) 1E 161348$-$5055 (hereafter 1E 1613) was discovered with {\em Einstein} \citep{tg1980}. It is located close to the center of the supernova remnant (SNR) RCW 103 with an age $\sim 2000-4000$ yr \citep{cc1976,npg1984,gph1997,cdb1997,cra2005}.
\citet{dcmtb2006} reported a $6.67$ hr periodicity with {\em XMM-Newton} observation, which may originate from either an orbital period of a low-mass X-ray binary (LMXB) or a spin period of an isolated neutron star (NS).
However, there is no counterpart in optical or radio detected \citep{tgmd1983,dmz2008},  casting doubt on the binary nature.
\citet{tkas2017} report the detection of an infrared counterpart with {\em Hubble Space Telescope} after
the latest outburst. The counterpart properties rule out the binary scenario and mimic
the infrared emission of isolated NSs.
%Recent observations of magnetar-like burst and outburst from 1E 1613 \citep{deb2016,rbe2016,bce2018} clearly indicate this source to be a magnetar with magnetic field of order $10^{15}$ G \citep{dcmtb2006,re2011}.
Observations of outbursts from 1E 1613 and the evolution of the spectral and timing properties along the outburst decay indicate this source to be a magnetar \citep{re2011,deb2016,rbe2016,bce2018}.

A problem in the magnetar scenario is how to spin down the NS from a birth period less than 1 s to about $6.67$ hr via magneto-dipole radiation within a few kyr \citep{dcmtb2006}.
Therefore, additional spin-down torque is required, possibly provided by a SN fallback disk, which interacted with the NS in the past propeller stage \citep{dcmtb2006,lxd2007}. Simulation of the spin evolution implies that the magnetic field could be of order $10^{15}$ G \citep{dcmtb2006}.
In this case the initial mass of the fallback disk plays a vital role for the following reasons.
(1) It provides useful information about the SN's explosion energy and the angular momentum transport in the ejected material \citep{perna2014}.
(2) It largely determines the evolution of the mass flow rate in the disk \citep{clg1990}. (3) The radiation of the disk in optical and infrared also depends on the initial disk mass, which  sets constraint on the possible detection limit in these wavelengths \citep{mh2001}.
However, estimates of the fallback disk mass in previous works are controversial.
\citet{dcmtb2006} suggested that a $5 \times 10^{15}$ G magnetar with a disk of mass $3\times 10^{-5} M_{\odot}$ could slow down from $300$ ms to $6.67$ hr over $2000$ yr.
\citet{ha2017} investigated the spin-down processes of 1E 1613 during the ejector and propeller stage. Assuming a constant mass flow rate ($\sim 10^{-12} M_\odot {\rm yr}^{-1}$), they showed that a NS with magnetic field of $5\times 10^{15}$ G can spin down from milliseconds to $6.67$ hr over its a few kyr lifetime. And the initial mass of the fallback disk was estimated to be $\sim 10^{-9} M_\odot$.
\citet{twlx2016} used a similar model but considered the evolution of the mass flow rate in the disk, and predicted a disk mass of $10^{-5} M_\odot$ with a comparable magnetic field.

In this work, we perform Monte-Carlo simulations of the spin evolution of a magnetar interacting with a fallback disk. Our method of calculating the spin-down torque is similar to that in previous works, but we take into account some important factors
(e.g., the conditions determining the inner and outer radii of the disk, and possible mass loss from the disk),
which was not included in those works.
%of the evolution of the fallback disk from super-Eddington accretion to be advection-dominated, and show that the passivity of the disk as well as the choice of the inner edge of the disk can strongly influence the efficiency of the slow-down processes.
We set up the model in section 2 and present the numerically calculated results in section 3. We summarize the results and discuss their implications in section 4.

\section{The Fallback Disk Model}

After the SN explosion, part of the ejected material may be gravitationally captured by the newborn NS  \citep{wse1993,chn2000,ama2001,wck2006,be2016}. If the fallback material possesses sufficient angular momentum, a disk may form around the NS\footnote{The disk is likely to be originally in the form of a ring.}. Similar as in the cases of tidal disruption events \citep{sm2012} and accretion-induced collapse \citep{ll2015}, the gravitationally captured fossil disk evolves following the viscous diffusion equation. If the viscosity coefficient $\nu$ depends on the radius $R$ in a power law, $\nu\propto R^n$, one can obtain self-similar solutions for the disk evolution \citep{pje1981,clg1990}.

The disk usually evolves through four phases with decreasing mass flow rate and expanding size as summarized in \citet{ll2015}: In phase 1, the newly formed disk takes the form of a slim or thick disk with a super-Eddington mass flow rate, characterized by advective cooling \citep{cg2009}. Phase 1 ends  when the outermost region of the disk is dominated by radiative cooling and turns to be geometrically thin.
In Phase 2, the disk comprises an inner slim disk surrounded by an outer thin disk.
Along with the decreasing mass flow rate, the outer thin disk region develops inwards.
When the inner slim disk region disappears, phase 2 ends and
the disk evolves to phase 3, which is geometrically thin and optically thick with efficient radiative cooling.
When the mass flow rate reduces to a few percent of the Eddington accretion rate, advection-dominated accretion flow (ADAF) appears in the inner region and phase 4 begins. Now the disk contains an outer thin disk region and an inner ADAF region developing outwards with decreasing mass flow rate. Radiative cooling is inhibited in the ADAF region by the low density of the gas, which leads to the heat kept in the region mostly advected onto the central object. The duration of different phases depends on the initial mass flow rate and the NS parameters (see Appendix).

%along with the decrease of the mass transfer rate and expansion of the disk.
%from slim or thick disk to advection-dominated accretion flow (ADAF)

%But in which conditions the fallback disk can be formed? We will talk about it following in this section.
%\subsection{Parameters}
%\subsubsection{Constant Parameters}

We first define several important radii that determine the state of a magnetized NS surrounded by a disk. The first is the magnetospheric radius which is taken to be the inner radius of the disk
\begin{equation}
R_{\rm in} = \xi R_{\rm A},
\end{equation}
where $\xi \sim 0.5-1$ \citep{gl1979,wym1996}, and $R_{\rm A}$ is the traditional Alfv\'en radius for spherical accretion
\begin{equation}
R_{\rm A} = \left(\frac{\mu^4}{2GM\dot{M}_{\rm in}^2} \right)^{1/7},
\end{equation}
where $\mu = BR_{\rm NS}^3$ is the magnetic dipole moment of the NS ($B$ and $R_{\rm NS}$ are the dipolar magnetic field and the radius of the NS, respectively), $G$ the gravitational constant, $M$ the NS's mass, and $\dot{M}_{\rm in}$ the mass flow rate at the inner edge of the disk.

The second is the corotation radius $R_{\rm c}$ where the local Keplerian angular velocity $\Omega_{\rm K}(R)$ equals the angular velocity $\Omega_{\rm s}$ of the NS,
\begin{equation}
R_{\rm c} = \left(\frac{GM}{\Omega_{\rm s}^2} \right)^{1/3}.
\end{equation}

The third is the light cylinder radius $R_{\rm lc}$ where the corotation velocity equals the speed of light $c$,
\begin{equation}
R_{\rm lc} = \frac{c}{\Omega_{\rm s}}=\frac{cP}{2\pi},
\end{equation}
where $P=2\pi/\Omega_{\rm s}$ is the spin period of the NS.

We assume that accretion onto the NS is allowed when $R_{\rm in} < R_{\rm c}$. Mass accretion also transfers angular momentum to the NS at a rate
\begin{equation}
N_{\rm in} = \dot{M}_{\rm acc}(GMR_{\rm in})^{1/2},
\end{equation}
where $\dot{M}_{\rm acc}$ is the accretion rate onto the NS.
Note that according to the magnetically threading disk model \citep{gl1979,wym1987,wym1995}, the interaction between the disk and the twisted magnetic field lines can also exert a torque on the NS. Since it is highly uncertain how efficient the magnetic torque can be, here we only take the simple form (5) for the accretion torque.
%Because the duration of the accretor phase is rather short (\textbf{usually less than tens of years}), the final result is not sensitive to the detailed form of the accretion torque.

When $R_{\rm c} < R_{\rm in} < R_{\rm lc}$, the NS is in the propeller phase. If we assume that all the material accreting onto the NS is ejected at the inner edge of the disk, the NS experiences a spin-down torque given by
\begin{equation}
N_{\rm prop}=-\dot{M}_{\rm in}R_{\rm in}v_{\rm esc}(R_{\rm in})\simeq-\dot{M}_{\rm in}(GMR_{\rm in})^{1/2}.
\end{equation}
where $v_{\rm esc}(R_{\rm in})$ is the escape velocity at the inner radius of the disk.
%Note that $\dot{M}_{\rm in}$ and $\dot{M}_{\rm acc}$ are not equivalent. $\dot{M}_{\rm acc}$ is the accreted mass onto the NS surface in unit time while $\dot{M}_{\rm in}$ is the flowing mass in the inner radius of the disk in unit time.

When $R_{\rm in} > R_{\rm lc}$, the NS is in the ejector phase and spins down only due to magnetic dipole radiation. The torque is given by
\begin{equation}
N_{\rm B}=-\frac{2\mu^2 \Omega^3}{3c^3}.
\end{equation}
We assume that $N_{\rm B}$ is also exerted on the NS in the accretor and propeller phases.

\subsection{The Effective Interaction Conditions}

The initial parameters of our fallback disk model include the initial spin period $P_0$ and the magnetic field $B$ of the NS (taken to be constant), the initial mass $M_{\rm d}$ and the initial outer radius $R_{\rm f}$ of the fallback disk.
The condition for the fallback disk formation is determined by comparing $R_{\rm in}$ with $R_{\rm f}$ at the disk formation time $t_{\rm f}$. The former represents the locatioin where the disk is disrupted by the NS magnetic field, and the latter is determined by the initial angular momentum of the fallback material.
If $R_{\rm in} < R_{\rm f}$ then the disk can develop. Otherwise, the disk is destroyed by the magnetic field and the NS evolves as an ejector.

In early works the disk formation time $t_{\rm f}$ was taken to be the dynamical timescale \citep[e.g.,][]{chn2000}.
\begin{equation}
%t_d \simeq 6.6 \times 10^{-5} T_{c,6}^{-1} R_{0,8}^{1/2} {\rm yr},
t_{\rm d} \simeq 0.05 {\rm \ s} \cdot R_{\rm f,8}^{3/2},
\end{equation}
where $R_{\rm f,8}=R_{\rm f}/10^8$ cm. Numerical calculations show that the self-similar evolution starts in a viscous timescale in the disk \citep{pje1991}.
\citet{clg1990} and \citet{mh2001} estimated the viscous time by use of the central temperature $T_{\rm c}$ of the disk
\begin{equation}
t_{\rm v1} \simeq 2.08 \times 10^{3} {\rm \ s} \cdot T_{\rm c,6}^{-1} R_{\rm f,8}^{1/2},
\end{equation}
where $T_{\rm c,6}=T_{\rm c}/10^6$ K,
while \citet{eeea2009} presented the estimate in terms of the initial disk mass $M_{\rm d}$
\begin{equation}
%t_v \simeq 3.19 \times 10^{-4} (M_{\rm d}/10^{-4}M_{\odot})^{-3/7} R_{0,8}^{25/14} {\rm yr},
t_{\rm v2} \simeq 1.0 \times 10^{4} {\rm \ s} M_{\rm d, -4}^{-3/7} R_{\rm f,8}^{25/14}.
\end{equation}
where $M_{\rm d, -4}=M_{\rm d}/10^{-4}M_{\odot}$. Generally $t_{\rm d}\ll t_{\rm v1}$ (unless $R_{\rm f}\gtrsim 4\times 10^{12}$ cm) and $t_{\rm v2}$.
Similar as \citet{fl2013}, we take the formation time $t_{\rm f}$ to be the maximum of $t_{\rm d}$, $t_{\rm v1}$ (assuming $T_{\rm c,6}=1$) and $t_{\rm v2}$.

\subsection{Disk Evolution}

\subsubsection{The mass flow rate and the accretion rate}

After the fallback disk is formed, its physical parameters vary in a self-similar way \citep{ll2015}
\begin{equation} \label{eq:12}
\begin{split}
& \dot{m}=\dot{m}_0 (t/t_{\rm f})^{-a}, \\
& r_{\rm out}=r_{\rm f} (t/t_{\rm f})^{2/3}, \\
& T=T_0 (\dot{m}/\dot{m}_0)^{1/4}, \\
& \rho=\rho_0 (\dot{m}/\dot{m}_0).
\end{split}
\end{equation}
Here $r_{\rm out}$ is the outer radius $R_{\rm out}$ of the disk  in units of the Schwartzschild radius $R_{\rm S} (\approx 4 \times 10^5$ cm for a $1.4 M_\sun$ NS), $r_{\rm f}=R_{\rm f}/R_{\rm S}$, and $\dot{m}$ is the mass flow rate $\dot{M}$ at $R_{\rm f}$ in units of the Eddington accretion rate $\dot{M}_{\rm Edd}$ ($ \approx 2\times 10^{18}$ gs$^{-1}$ for a $1.4M_\odot$ NS with radius of $10^6$ cm).
The index $a$ equals $4/3$ or 1.25 depending on opacity \citep{clg1990,fwb2002}, $T$ and  $\rho$ are the temperature and the density of the disk material, respectively. The subscript 0 denotes the values evaluated at $t=t_{\rm f}$. From  \citet{cg2009} we have
\begin{equation} \label{eq:12}
\begin{split}
& T_0 \simeq (0.92 \times 10^8 {\rm \ K}) \dot{m}_0^{1/4} r_{\rm f}^{-5/8}, \\
& \rho_0 \simeq (2.14 \times 10^{-4} {\rm \ g\,cm}^{-3}) \dot{m}_0 r_{\rm f}^{-3/2}, \\
& \dot{m}_0=\frac{M_{\rm d}}{\dot{M}_{\rm Edd} t_{\rm f}},
\end{split}
\end{equation}
where $M_{\rm d}$ is the initial disk mass.

We assume that if the mass flow rate $\dot{M}$ is higher than a critical rate\footnote{The value of $\dot{M}_{\rm cr}$ is derived for spherical accretion onto a NS by equating the shock radius with the photon trapping radius \citep{cra1989}. For disk accretion this value might change with the outer radius of the disk and we adopt the same value for simplicity.} $\dot{M}_{\rm cr} \simeq 1.9 \times 10^{22}$ g s$^{-1}$, neutrino loss rather radiation dominates the accretion flow and all the mass transferred is accreted by the NS \citep{csa1971,cra1989}. When $\dot{M}$ is less than $\dot{M}_{\rm cr}$ but higher than the local Eddington accretion rate, then the Eddington-limited accretion is enabled and there is mass loss from the disk starting at the spherization radius\footnote{The spherization radius is the radius in the disk where the {\em local} accretion luminosity exceeds the Eddington luminosity.} \citep{ss1973,lgv1999,g2017}
\begin{equation}
R_{\rm sph}=\frac{3}{2}\frac{GM\dot{M}}{L_{\rm Edd}}.
\end{equation}
The mass flow rate at the inner edge of the disk is
\begin{equation}
\dot{M}_{\rm in}=
\left\{
\begin{split}
& \dot{M}, \ \ \ \ \ \  \ \ \ \ \ \ \ \ \ \ {\rm if}\	\dot{M}>\dot{M}_{\rm cr}\ {\rm or\ if}\ \dot{M}<\dot{M}_{\rm cr}\ {\rm and}\ R_{\rm in} > R_{\rm sph} \\
& \dot{M}(R_{\rm in}/R_{\rm sph}), \ \ {\rm if}\ \dot{M}<\dot{M}_{\rm cr}\ {\rm and}\ R_{\rm in}\leq R_{\rm sph}. \\
\end{split}
\right.
\end{equation}
Note that when $R_{\rm in}< R_{\rm sph}$, combination of Eqs.~(1), (2), and (14) leads to a new expression of $R_{\rm in}$
\begin{equation}
R_{\rm in}=\xi^{7/9} \left(\frac{\mu^4}{2GM}\right)^{1/9} \left(\frac{2L_{\rm Edd}}{3GM}\right)^{-2/9}
\end{equation}
since $\dot{M}_{\rm in}=2L_{\rm Edd}R_{\rm in}/(3GM)\simeq 0.68(\xi L_{\rm Edd})^{7/9}\mu^{4/9}(GM)^{-8/9} $ does not vary with $\dot{M}$.

%The accretion rate of the NS is then
%\begin{equation} \label{eq:12}
%\dot{M}_{\rm acc}=
%\left\{
%\begin{split}
%& \dot{M}_{\rm in}, \ \ \ \ \ \	{\rm if} \ \dot{M}_{\rm in}\leq\dot{M}_{\rm Edd} \\
%& \dot{M}_{\rm Edd}, \ \ \ \ {\rm if} \ \dot{M}_{\rm in} > \dot{M}_{\rm Edd}. \\
%\end{split}
%\right.
%\end{equation}

\subsubsection{Conditions at the outer disk radius}

The disk expands due to viscous dissipation. We specify that its outer radius increases from $R_{\rm f}$ at the time $t_{\rm f}$ to $R_{\rm out}$ at a time $t(>t_{\rm f})$ if viscous dissipation always works.
However, there are two conditions that may further constrain the outer radius of the disk. For a differentially rotating disk there exists a so-called self-gravity radius $R_{\rm sg}$ beyond which self-gravity is important and the disk becomes unstable \citep{lp1974,pje1991}. In our case it can be estimated to be  \citep{bks1998}
\begin{equation}
R_{\rm sg} \simeq 9.62 \times 10^{10} \ {\rm cm} \cdot \left(\frac{\rho}{1 \ {\rm g \ cm}^{-3}}\right)^{-1/3}.
\end{equation}
Meanwhile, the temperature in the disk decrease with radius, so the outermost disk will become neutral eventually \citep{mh2001,eeea2009} if the temperature drops to a critical value (around $300$ K) \citep{ama2001}. So for a given mass flow rate there exists a so-called neutralization radius $R_{\rm neu}$, beyond which the disk becomes passive and there is hardly any transport of mass or angular momentum.  Since the energy dissipation rate in the disk reduces with decreasing mass flow rate, $R_{\rm neu}$ can be expressed as \citep{ll2015}
\begin{equation}
R_{\rm neu} = R_{\rm f} \cdot  \left(\frac{T_4}{300 \ {\rm K}}\right)^{4/3} \cdot \left(\frac{k_5}{\dot{m}_0}\right)^{2/5} \cdot \left(\frac{t}{t_{\rm f}}\right)^{-19/35}.
\end{equation}
where the parameters $T_4$ and $k_5$ are described in Appendix.

Therefore, the actual outer radius $R_{\rm out, act}$ of an active disk should take the smallest value among $R_{\rm out}$, $R_{\rm sg}$, and $R_{\rm neu}$, i.e., $R_{\rm out, act}=\min(R_{\rm out},R_{\rm sg},R_{\rm neu})$. The existence of an active disk further requires
$R_{\rm in}<R_{\rm out, act}$. Otherwise the NS spins down only via magnetic dipole radiation.

We summarize the conditions for the working interaction between the fallback disk and the NS as follows:
\begin{enumerate}
	\item the formation time $t_{\rm f}$  of the fallback disk is less than the age of 1E 1613;
	%\item the inner radius $R_{\rm in}$ is smaller than the outer radius $R_{\rm f}$ at $t_{\rm f}$;
	\item the magnetospheric radius $R_{\rm in}$ at $t_{\rm f}$ is smaller than the initial outer radius $R_{\rm f}$;
	\item the disk keeps viscously active.
\end{enumerate}

\section{Results}

\subsection{The reference model}

We first set up the parameters for the reference model in our Monte-Carlo simulation: $a=4/3$, $\xi=0.5$, $B=10^{15}$ G, $P_0=0.01$ s, and
$M_{\rm d}$ and $R_{\rm f}$ are randomly distributed in the range of $[10^{-10},10^{-1}]$\,$M_\odot$ and $[R_{\rm NS},10^6 R_{\rm S}]\,$, respectively.
We simulate the evolution for a sample of $10^6$ NSs.

Figure 1 shows the distribution of the formation time $t_{\rm f}$ of the fallback disks against their initial mass $M_{\rm d}$ (left panel) and initial outer radius $R_{\rm f}$ (right panel), where the blue and red colors mean that the disks can and cannot form, respectively. The condition for disk formation is that the inner disk radius at the time $t_f$, which depends on both the NS magnetic field and the mass flow rate, should be less than $R_{\rm f}$. Otherwise the disk is assumed to be disrupted by the magnetic field and does not influence the subsequent evolution of the NS.
We use the color depth to describe the number percentage of NSs in a specific parameter space among the total sample.
It is seen that the disk formation requires relatively large initial mass or outer radius.
%This can be understood as follows.
%Combining Eqs.~(1), (2), (10), and (12), we have $R_{\rm mag}\propto \dot{m}_0^{-2/7}\propto (M_{\rm d}/t_{\rm f})^{-2/7}\propto M_{\rm d}^{-20/49}R_{\rm f}^{25/49}$. Then from the condition for disk formation $R_{\rm mag}< R_{\rm f}$, wen can get $R_{\rm f}>M_{\rm d}^{-5/6}$.
The black solid line in Fig.~1 represents the possible age of RCW 103 (3200 yr), which is taken to be the ending time in our calculations. So the parameters resulting in $t_{\rm f}$ longer than 3200 yr are excluded in our simulation.

%To avoid the singularity problem in Eq.~(10) when $t=0$, we can either set the starting time to be a very small value like $t_0 = 0.0001$ s or rewrite the term $(t/t_{\rm f})$  to be $(1+t/t_{\rm f})$.
%We compare the spin evolutions in Fig.~2 in the two cases adopting $M_{\rm d} = 10^{-5} M_{\odot}$ and $R_{\rm f} = 10^4 R_{\rm S}$.
%The red line represents the period of 1E1613 with age from $2000$ to $4000$ yr.
%The whole evolution is divided into two parts, with the dashed and solid lines representing the evolution before and after the disk formation, respectively.
%There is very little difference between the results.

Figure 2 shows the distribution of the final spin period $P$ versus $M_{\rm d}$ (left panel) and $R_{\rm f}$ (right panel) in the reference model. The blue region demonstrates the case with disk interaction, and the red line (corresponding to a period of about 14 s) indicates the case without disk interaction (either no disk forms or the disk is passive).
One can get the following information from the figure. (1) When $M_{\rm d}\sim 10^{-1} M_{\odot}- 10^{-6} M_{\odot}$, the spin period  increases with decreasing $M_{\rm d}$. The reason is that, with such a massive disk, the NS's spin period is able to reach the equilibrium spin period at 3200 yr \citep[see also][]{chn2000}
\begin{equation}
P_{\rm eq} = 2^{11/14} \pi (G M)^{-5/7} \xi^{3/2} \mu^{6/7} \dot{M}_{\rm in}^{-3/7}.
\end{equation}
Combining Eqs.~(11), (12), and (18), we have $P_{\rm eq} \propto \dot{M}_{\rm in}^{-3/7}\propto M_{\rm d}^{-18/49}$.
(2) When $M_{\rm d}<\sim 10^{-7}-10^{-6} M_{\odot}$,
the NS-disk interaction is unable to decelerate the NS to the equilibrium spin period, and the final period roughly decreases with decreasing $M_{\rm d}$. The reason is that a smaller disk results in a smaller spin-down torque.
(3) The maximum period is $\sim 2000$ s with $M_{\rm d} \sim 10^{-7} - 10^{-4} M_{\odot}$. This results from a combination of the following two requirements:  $M_{\rm d}$ should be large enough to ensure that the NS-disk interaction can lead the NS to reach the equilibrium spin period, and  $M_{\rm d}$ should be small enough to ensure sufficiently low mass accretion rate and hence long equilibrium spin period.
We also note that the final spin period  is not sensitively dependent on the initial outer radius $R_{\rm f}$, which ranges from $10^2  R_{\rm S}$ to $10^5 R_{\rm S}$.

\subsection{Parameter study}

We have found that in the reference model the maximum spin period is significantly shorter than that of 1E 1613. In this subsection, we vary the values of the input parameters ($P_0$, $B$, $a$ and $\xi$) based on the reference model, to see how they can influence the spin evolution of of the NS.

\subsubsection{The initial spin period $P_0$}
We first change the initial spin period of the NS. The calculated spin periods are shown in Fig.~3 against $M_{\rm d}$ (left panels) and $R_{\rm f}$ (right panels). From top to bottom, the initial spin period is taken to be 0.001, 0.01, 0.1, and 1 s. The meanings of the symbols are same as in Fig.~2.
It is seen that the maximum period $P_{\rm max}$ in each panel is similar, around 2000 s.
Figure 4 demonstrate the spin evolution with $M_{\rm d} = 10^{-5} M_\odot$ and $R_{\rm f} = 10^4 R_{\rm S}$. We also add two cases with the initial spin period of 10 and 100 s.
In each case the solid and dashed curves represent the spin evolution stages with and without NS-disk interaction, respectively.
%Except the case of  $P_0 = 1$ s, the spin period evolutions converge when $t>\sim 10^8$ s. The reason is that the magnetic dipole radiation torque, $N_B \propto P^{-3}$, is larger for smaller $P_0$, leading to a faster spin-down of the NS. So before the disk starts to work, the NS has already evolved to the similar periods. When $P_0 = 1$ s, the spin period changes little before the disk is formed, but the subsequent propeller torque can efficiently spin down the NS to the equilibrium period.
The six curves coincide at the time $\sim 10^9$ s, when the NS has reached the equilibrium spin period by the propeller torque.
Thus we conclude that the final spin period is insensitive to the choice of $P_0$.

\subsubsection{The initial magnetic field $B$}
Figure~5 shows the final spin period distribution with different initial magnetic field $B$.  From top to bottom, we take $B = 10^{14}$, $5 \times 10^{14}$, $10^{15}$, and $5 \times 10^{15}$ G. Other parameters are same as in the reference model. Obviously $P_{\rm max}$ becomes larger for stronger $B$, which is  75 s,  838 s,  2368 s, and 15740 s, respectively. This relation closely reflects that the NS has reached the equilibrium spin.
The red lines show the spin period evolution without disk interaction. They vary with the adopted magnetic field strength since the spin-down torque $N_{\rm B}\propto B^2$.
Figure~6 compares the spin evolutions with different values of the magnetic field, which confirms the results obtained by the Monte-Carlo calculation. A remarkable feature is that the maximum periods are considerably less than the spin period of 1E 1613, even if a very high magnetic field of $5 \times 10^{15}$ G is adopted.

\subsubsection{The power index $a$}
The final spin period is also determined by the mass flow rate in the fallback disk. As shown in Eq.~(11), $\dot{M}_{\rm in}$ depends on a power law with the index $a$. Given $\dot{m}_0$ and $t_{\rm f}$, the larger $a$, the smaller $\dot{M}$, hence the longer $P_{\rm eq}$. However, a smaller $\dot{M}$ means that the disk could become passive and stop the interaction with the NS at an earlier time.
%From Eq.~(10) and (11), one can get $\dot{M} = M_{\rm d} t^{-a} t_{\rm f}^{a-1}$.
The value of $a$ depends on the opacity law in the disk, ranging between $19/16$ and $4/3$ \citep{clg1990,pje1991,cg2009}. In the reference model the power index $a$ in Eq.~(11) is taken to be $4/3$. Now we use another value of 1.25 and show the calculated results in Fig.~7.
The top and bottom panels compare the results with these two values of $a$.
There is no remarkable difference between the two cases.
The maximum spin periods are both around 2000 s.

\subsubsection{The parameter $\xi$}
The equilibrium period $P_{eq} \propto \xi^{3/2}$, indicating that $\xi$ may also be an important parameter. It relates the inner disk radius with the Alf\'ven radius for spherical accretion. However, the value of $\xi$ is poorly known. It is usually taken to be 0.5 \citep{gl1979,long2005}, but \citet{wym1996} argued that $\xi\simeq 1$ may be a better choice.
We compare the spin period distribution with $\xi=1$ and 0.5 in Fig.~8. The maximum spin period increases to be $\sim 6000$ s when $\xi=1$, which is about 3 times of that in the reference model.
%Fig.~11 shows a detailed example evolution

\subsubsection{Parameter space for the $6.67$ hr final period}

From the above analysis we find that the parameters that can significantly influence $P_{\rm max}$ are $M_{\rm d}$, $R_{\rm f}$, $B$, and $\xi$. Larger $B$ and $\xi$ lead to longer $P_{\rm max}$,
so we fix $B=5 \times 10^{15}$ G and $\xi=1$ and redo the Monte-Carlo simulation to search for the parameter space that can account for the spin period of 1E 1613. The results are shown in Fig.~9, where the top and bottom panels correspond to the age of 3200 yr and 4000 yr, respectively.
In the former case we get $P_{\rm max} \simeq 6.5$ hr with $M_{\rm d}\sim 8\times 10^{-8} M_\odot$ and $R_{\rm f}\sim 3\times 10^5 R_{\rm S}$. In the latter we get $P_{\rm max} \simeq 7.6$ hr with $M_{\rm d}\sim 6\times 10^{-8} M_\odot$ and $R_{\rm f}\sim 3\times 10^5 R_{\rm S}$, and $P \gtrsim 6.67$ hr with $M_{\rm d}\sim (6 - 14) \times 10^{-8} M_\odot$ and $R_{\rm f}\sim (1.8 - 3.6)\times 10^5 R_{\rm S}$. The very small parameter space suggests that 1E 1613 is a rare object.
In Fig.~10 we show the spin evolutionary track with the initial parameters taken to be $B=5 \times 10^{15}$ G, $P_0 = 0.01$ s, $\xi=1$, $M_{\rm d}=8.9 \times 10^{-8} M_{\odot}$, and $R_{\rm f} =3.5 \times 10^5 R_{\rm S}$.
The final spin period at the age of 4000 yr is 24042 s.
The solid and dashed curves represent the spin evolutionary stages with and without NS-disk interaction, respectively. Most of the spin-down is caused by the propeller torque exerted by the disk.
%Having reached the spin equilibrium, the NS is currently spinning down at a rat  $\sim 4$ s yr$^{-1}$ from our calculation.
%The right panel shows the results of different parameters: the blue line with $B= 10^{15}$ G, the green line with $\xi=0.5$, the yellow line with $P_0 = 1$ s and the carmine line with $M_{\rm d}=6.31 \times 10^{-8} M_{\odot}$ and $R_{\rm f} =3.47 \times 10^5 R_{\rm S}$
%while other parameters are same with left panel, respectively.

\section{Discussion}
In this work we investigate how a fallback disk can help the NS 1E 1613 spin down to the $6.67$ hr period within about 3200 yr.
We show that if 1E 1613 was born with $B\gtrsim 5\times 10^{15}$ G, it can be spun down to extremely slow spin by the NS-fallback disk interaction, provided that the initial mass and outer radius of the disk are around $\sim 10^{-7} M_\odot$ and $3\times 10^5\,R_{\rm S}$, respectively.
%We show that if 1E 1613 was born as a magnetar, it is possible to account for its \textbf{super slowly spinning} provided that its magnetic field strength $\gtrsim 5\times 10^{15}$ G and the initial fallback mass $\sim 10^{-7} M_\odot$, and only a very small parameters space meets the requirement.
This implies that 1E 1613 is indeed rare. More generally, when $M_{\rm d}<\sim 10^{-7} M_\odot$, the NS is likely to be in the ejector phase and acts as a pulsar or a normal magnetar. With a more massive fallback disk the NS experiences less efficient spin-down because of higher mass flow rate. A comparison of the spin evolution with $M_{\rm d}=10^{-9}$, $10^{-7}$, $10^{-5}$, and $10^{-3} M_\odot$ is shown in Fig.~11.

Our results share some similarities with previous studies \citep{ha2017,twlx2016}, but are in contradiction with them in some respects. In Fig.~12 we plot our simulated results in the $P$ vs. $M_{\rm d}$ plane. The left and right panel corresponding to $\xi=0.5$ and 1, respectively. The results of \citet{ha2017}  and \citet{twlx2016} are depicted with the red square in the left panel and the red circle in the right panel, respectively. It is obviously seen that they are outside of the predicted region in our simulation. The possible reasons for this discrepancy are as follows.
\citet{ha2017} used a small constant mass flow rate $\dot{M} = 10^{-12} M_{\odot} {\rm yr}^{-1}$ to calculate the NS spin evolution. With such a mass flow rate, we can infer that either $M_{\rm d}$ is extremely small so the disk is disrupted by the NS magnetic field at birth, or $t_{\rm f}$ is longer than the lifetime of 1E 1613. In both cases there is no efficient NS-disk interaction to spin down the NS.
\citet{twlx2016} considered an evolving mass flow rate, and got a required initial disk mass $\sim 10^{-5} M_{\odot}$ by assuming $t_{\rm f}=2000$ s. With such a disk mass, the viscous time in the disk is estimated to be $\sim 10^7 - 10^{11}$ s depending on the magnitude of $R_{\rm f}$. This will lead to a shorter propeller episode and hence a smaller spin period.
However, if we ignore the requirements on the disk formation, we can obtain comparable results with them. This can be seen that, the two symbols are confined by the two orange lines, which reflect the lower and upper boundaries of the equilibrium spin periods.

\section{Summary}

In this work we explore the possible parameter space in the fallback disk model that can reproduce the extremely slow spin
of 1E 1613. Compared with previous studies, we take into account some important factors that affect the formation and evolution of the fallback disk, including the conditions determining the inner and outer radii of the disk, and possible mass loss from the disk.
Our results show that an ultrastrong magnetic field $\gtrsim 10^{15}$ G  is not sufficient to explain its current spin period. Moreover, a fallback disk with initial mass $\sim 10^{-7} M_\sun$ is required. More or less massive disks are unable to spin down the NS to 6.67 hr period within $\sim 3-4$ kyr.  This implies that the current state of 1E1613 is actually very difficult to reach, so its evolution can provide very interesting constraints on the supernova fallback model. Among the known magnetars, 1E1613 is the only one with very long spin period. If future observations discover a population of extremely slowly spinning magnetars like 1E1613, it means that either the parameters of supernova fallback occupy a limited range or the model of fallback evolution should be significantly revised.

\acknowledgements We thank the referee for helpful comments. This work was supported by the National Key Research and Development Program of China (2016YFA0400803), the Natural Science Foundation of China under grant Nos. 11333004 and 11773015, Project U1838201 supported by NSFC and CAS.

\appendix

\section{The parameters $T_4$ and $k_5$}
From \citet{ll2015} we provide the derivation and expressions of $T_4$ and $k_5$ as follows.
\begin{equation} \label{eq:12}
\begin{split}
& t_1 = t_{\rm f} \left(\frac{\dot{m}_0}{r_{\rm f}}\right)^{1/2}, \\
& k_2 = r_{\rm f} \left(\frac{t_1}{t_{\rm f}}\right)^{38/21}, \\
& \rho_2 = \rho_0 \left(\frac{r_{\rm f}}{\dot{m}_0}\right)^3, \\
& T_2 = T_0 \left(\frac{t_1}{t_{\rm f}}\right)^{1/4} \left(\frac{r_{\rm f}}{\dot{m}_0}\right)^{1/4}, \\
& t_{gas} = t_{\rm f} \left[\frac{290}{r_{\rm f}} \cdot k_2^{16/21} \left(\frac{\dot{m}_0}{k_2}\right)^{4/9}  \left(\frac{r_{\rm f} \cdot \rho_2^{1/3} \cdot m^{2/3}}{2.92 \times 10^5}\right)^{2/3}\right]^{441/160}, \\
& T_3 = T_2 \left[290 \cdot \frac{\dot{m}_0 ^{16/21}}{r_{\rm f}}\right]^{3/20} \times \left(\frac{\dot{m}_0}{k_2}\right)^{2/7} \left(\frac{t_{gas}}{t_{\rm f}}\right)^{16/49}, \\
& T_4 = T_3 \left(1.5 \times 10^4 \cdot \frac{\dot{m}_0^{2/3}}{r_{\rm f}}\right)^{-3/20}, \\
& k_5 = k_2 \left(\frac{t_{gas}}{t_{\rm f}}\right)^{3/14}.
\end{split}
\end{equation}

\section{The time scale for the phases of disk evolution}
We present the time scale for each phase of the disk evolution as follows.
\begin{equation}
t_1 = t_{\rm f} \left(\frac{\dot{m}_0}{r_{\rm f}}\right)^{1/2} \approx
\left\{
\begin{split}
& 3.16 \times 10^{-3} {\rm \ s} \cdot \dot{m}_0^{1/2} R_{\rm f,8} ,
\ \ \ \ \ \ \ \ \ \ \ \ \ \ \ {\rm if}\	t_{\rm f} = t_{\rm d} \\
& 132 {\rm \ s} \cdot \dot{m}_0^{1/2} ,
\ \ \ \ \ \ \ \ \ \ \ \ \ \ \ \ \ \ \ \ \ \ \ \ \ \ \ \ \ \ \ {\rm if}\	t_{\rm f} = t_{\rm v1} \\
& 632 {\rm \ s} \cdot M_{\rm d, -4}^{-3/7} R_{\rm f,8}^{9/7} \dot{m}_0^{1/2},
\ \ \ \ \ \ \ \	\ \ \ \ \ \ \ \ \  {\rm if}\ t_{\rm f} = t_{\rm v2}. \\
\end{split}
\right.
\end{equation}
\begin{equation}
t_2 = t_{\rm f} \left(\frac{k_5}{r_{\rm in}}\right)^{14/19} \approx
\left\{
\begin{split}
& 1.28 \times 10^{-3} {\rm \ s} \cdot \xi^{-14/19} \mu_{30}^{-8/19} R_{\rm f,8}^{1424/969} \dot{m}_0^{712/969} \dot{m}_{\rm in}^{4/19}  ,
\ \ \ \ \ \ \ \ \ \ \ {\rm if}\	t_{\rm f} = t_{\rm d} \\
& 53 {\rm \ s} \cdot \xi^{-14/19} \mu_{30}^{-8/19} R_{\rm f,8}^{455/969} \dot{m}_0^{712/969} \dot{m}_{\rm in}^{4/19}  ,
\ \ \ \ \ \ \ \ \ \ \ \ \ \ \ \ \ \ \ \ \ \ \ \ {\rm if}\	t_{\rm f} = t_{\rm v1} \\
& 257 {\rm \ s} \cdot \xi^{-14/19} \mu_{30}^{-8/19} M_{\rm d,-4}^{-3/7}  R_{\rm f,8}^{11906/6783} \dot{m}_0^{712/969} \dot{m}_{\rm in}^{4/19} ,
\ \ \ \ \ \ \ \ \ \ \ {\rm if}\	t_{\rm f} = t_{\rm v2}. \\
\end{split}
\right.
\end{equation}
\begin{equation}
t_3 = t_{\rm f} (k_5 \cdot r_{\rm in}^2)^{14/19} \approx
\left\{
\begin{split}
& 468 {\rm \ s} \cdot \xi^{28/19} \mu_{30}^{16/19} R_{\rm f,8}^{1424/969}  \dot{m}_0^{712/969} \dot{m}_{\rm in}^{-8/19} ,
\ \ \ \ \ \ \ \ \ \ \ \ \ \ \ \ \ \ \ \ \ \ \ {\rm if}\	t_{\rm f} = t_{\rm d} \\
& 1.94 \times 10^7 {\rm \ s} \cdot \xi^{28/19} \mu_{30}^{16/19} R_{\rm f,8}^{155/969}  \dot{m}_0^{712/969} \dot{m}_{\rm in}^{-8/19} ,
\ \ \ \ \ \ \ \ \ \ \ \ \ \ \ \, {\rm if}\	t_{\rm f} = t_{\rm v1} \\
& 9.35 \times 10^7 {\rm \ s} \cdot \xi^{28/19} \mu_{30}^{16/19} M_{\rm d, -4}^{-3/7} R_{\rm f,8}^{11906/6783}  \dot{m}_0^{712/969} \dot{m}_{\rm in}^{-8/19}, \ \ \ \ {\rm if}\	t_{\rm f} = t_{\rm v2}. \\
\end{split}
\right.
\end{equation}
where $r_{\rm in} = R_{\rm in} / R_{\rm S}$, $\dot{m}_{\rm in} = \dot{M}_{\rm in} / \dot{M}_{\rm Edd}$ and $\mu_{30} = \mu /(10^{30} {\rm \ G \ cm^3})$, and $M_{\rm d, -4}=M_{\rm d}/{10^{-4}M_{\odot}}$.

For example, for the parameters of the reference model: $\xi = 0.5$, $a = 4/3$ and $M_{\rm d} = 10^{-5} M_\odot$, $R_{\rm f} = 10^4 R_{\rm S}$, $B = 10^{15}$ G, we get $t_{\rm f} \approx 0.62$ yr, $\dot{m}_0 \approx 514$ and $t_1 \approx 0.14$ yr, $t_2 \approx 0.5$ yr, $t_3 \approx 2.6 \times 10^{15}$ yr. For the parameters in Fig.~10: $\xi = 1$, $a = 4/3$ and $M_{\rm d} = 8.9 \times 10^{-8} M_\odot$, $R_{\rm f} = 3.5 \times 10^5 R_{\rm S}$, $B = 5 \times 10^{15}$ G, we get $t_{\rm f} \approx 2670$ yr, $\dot{m}_0 \approx 10^{-3}$ and $t_1 \approx 0.15$ yr, $t_2 \approx 0.07$ yr, $t_3 \approx 2.7 \times 10^{16}$ yr. In both cases the NS spends most of its lifetime in phase 3.

\newpage

%\begin{figure}
%	\centering
%	\figurenum{1}
%	\gridline{\fig{b15p001_xi05_nobury_mtf.eps}{0.3\textwidth}{(a)}
%	          \fig{b15p001_xi05_nobury_rftf.eps}{0.3\textwidth}{(b)}
%	          }
%	\caption{The formation time of a fallback disk.\label{fig:fig1}}
%\end{figure}

%\begin{figure}
%\centering
%\figurenum{1}
%\includegraphics[width=0.45\textwidth]{b15p001_xi05_nobury_mtf.eps}
%\includegraphics[width=0.45\textwidth]{b15p001_xi05_nobury_rftf.eps}
%\caption{The formation time of fallback disks. The blue points represent the parameters that the disk can be formed while the red points for the disk can't be formed. The black solid lines is the life time of 1E1613, which is the stop time of our code.}
%\end{figure}

\begin{figure}
%\figurenum{1}
\plotone{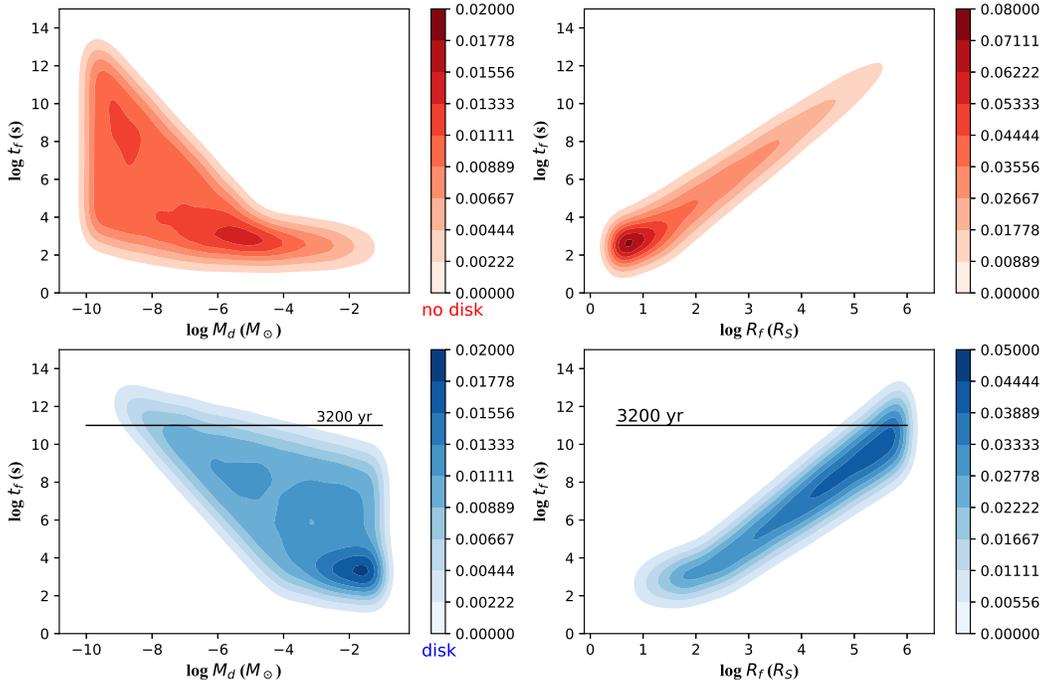}
\caption{Distribution of the formation time $t_{\rm f}$ of the fallback disks against the initial mass $M_{\rm d}$ (left panels) and the initial outer radius $R_{\rm f}$ (right panels). In the upper and lower panels we use the red and blue colors to represent the cases  that disk cannot and can form, respectively. The color depth describes the number density of NSs in a specific parameter space. The black solid line represents the possible age of RCW 103 (3200 yr), which is taken to be the ending time in our calculations.\label{fig:f2}}
\end{figure}

%\begin{figure}
%	\figurenum{2}
%	\plotone{fig2.eps}
%	\caption{Comparing the spin evolutions in the two cases to avoid the singularity problem in Eq.~(10) when $t=0$ adopting $M_{\rm d} = 10^{-5} M_{\odot}$ and $R_{\rm f} = 10^4 R_{\rm S}$, which are setting the starting time to be a very small value like $t_0 = 0.0001$ s (the black curve) or rewrite the term $(t/t_{\rm f})$  to be $(1+t/t_{\rm f})$ (the yellow curve). The whole evolution is divided into two parts, with the dashed and solid lines representing the evolution before and after the disk formation, respectively. The red line shows the spin period of 1E 1613 with age $2000-4000$ yr.\label{fig:f2}}
%\end{figure}

\begin{figure}
\centering
%\figurenum{2}
\plotone{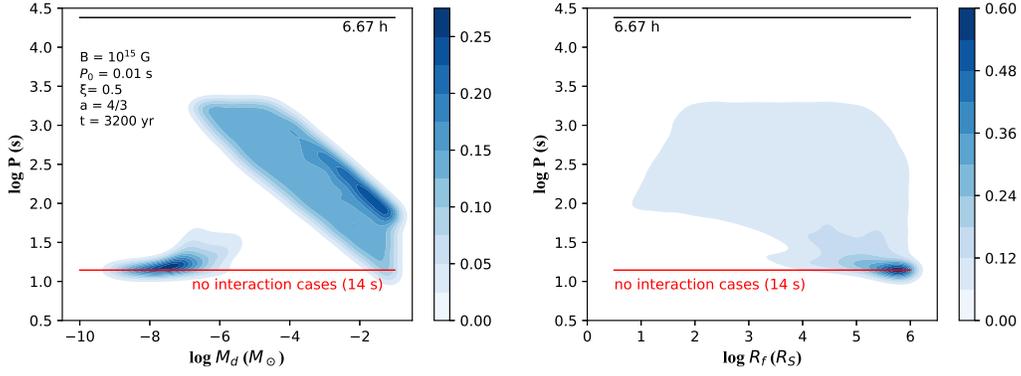}
\caption{Distribution of the final spin period against $M_{\rm d}$ (left panel) and $R_{\rm f}$ (right panel) in the reference model.
The values of the initial parameters are listed in the left panel. The black solid line represents the period (6.67 h) of 1E 1613.
The blue region shows the cases with disk interaction, and the red line (corresponding to a period of about 14 s) shows the cases without disk interaction because either no disk can form or the disk is passive.}
\end{figure}

\begin{figure}
	\centering
	%\figurenum{3}
	\includegraphics[width=0.9\textwidth]{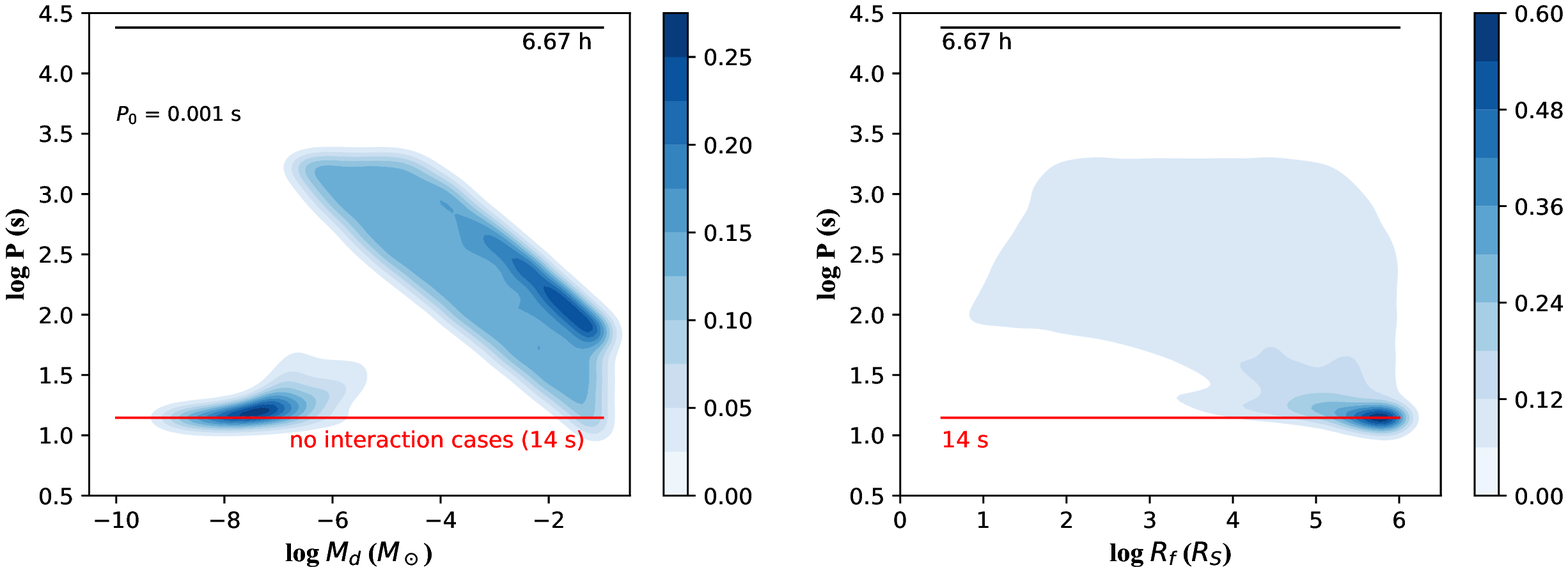}
	\includegraphics[width=0.9\textwidth]{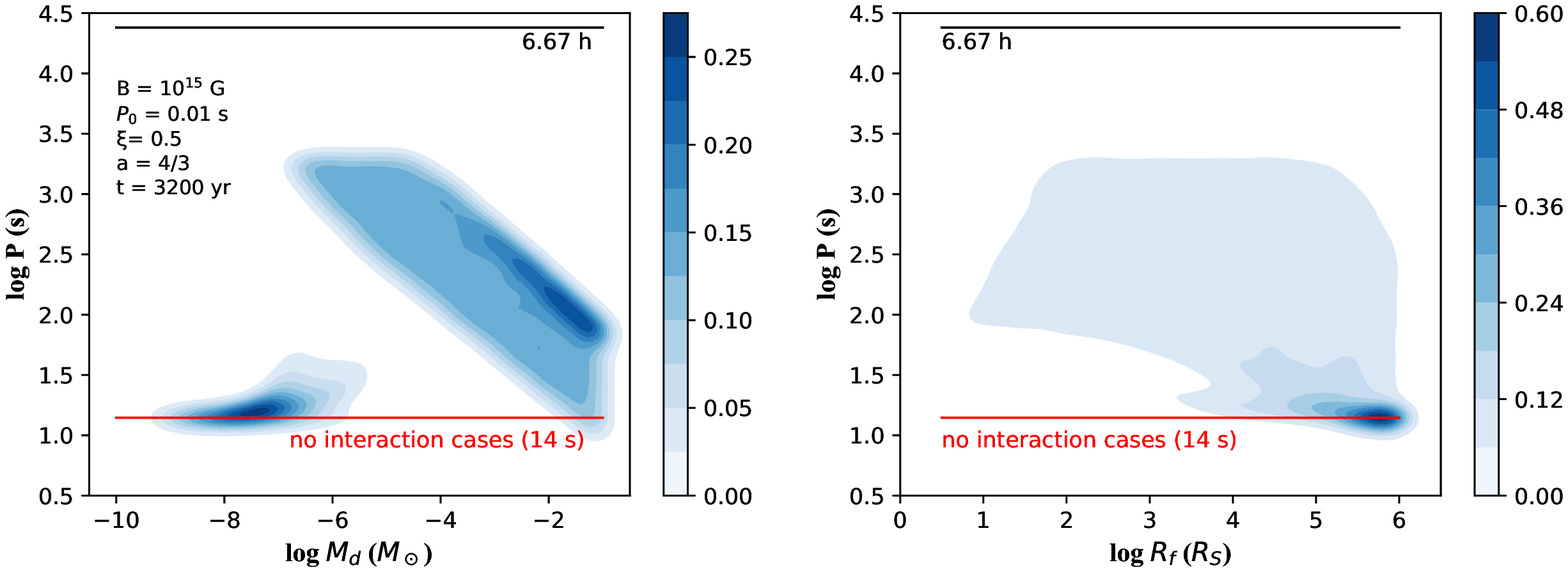}
	\includegraphics[width=0.9\textwidth]{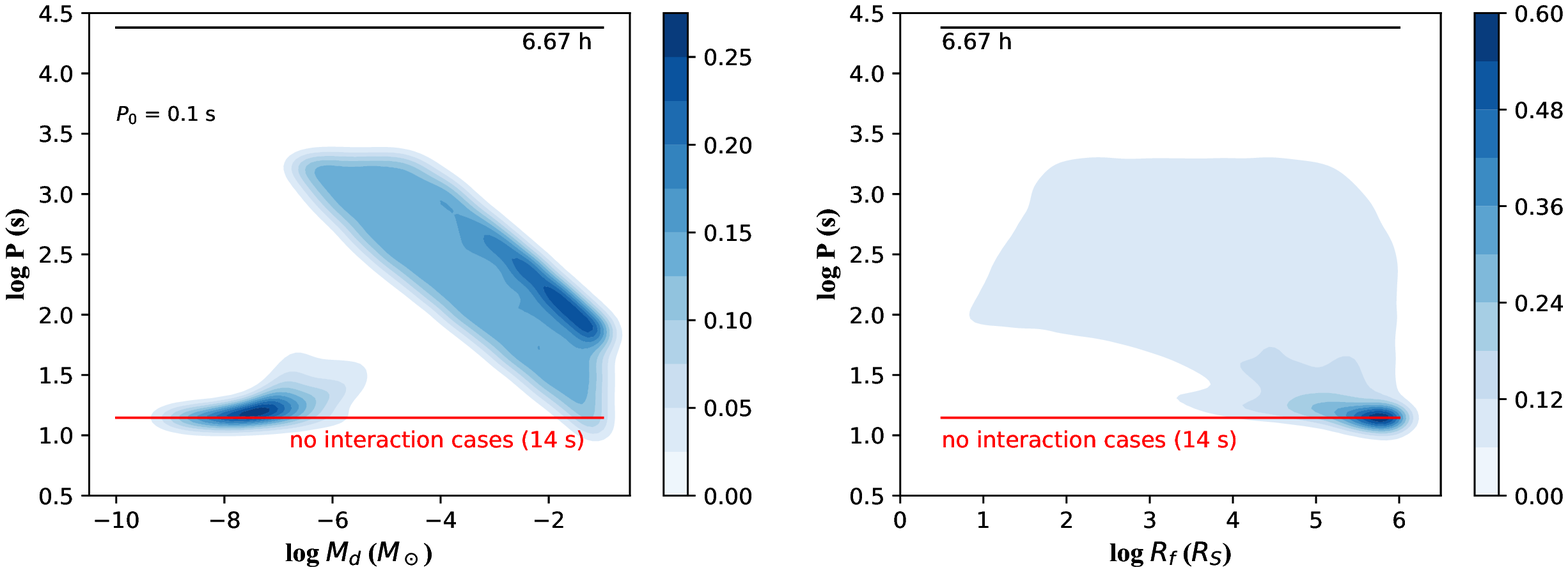}
	\includegraphics[width=0.9\textwidth]{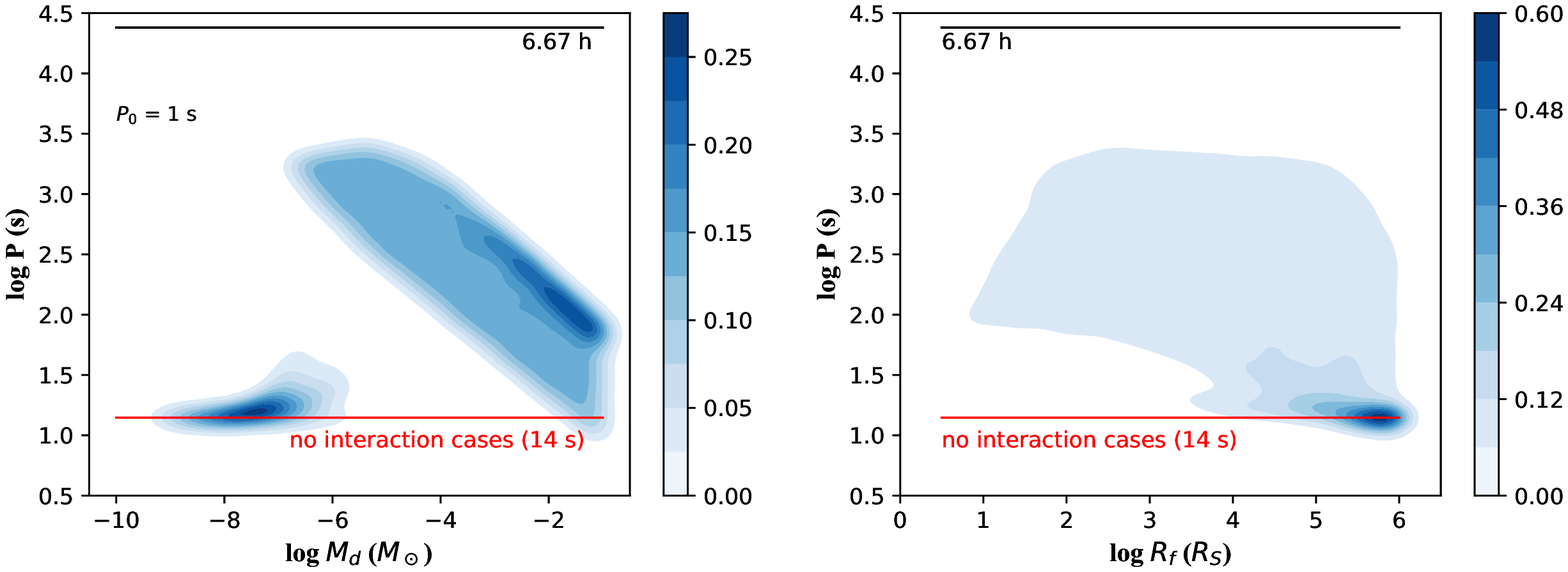}
	\caption{Same as Fig.~2, but with the initial spin period taken to be 0.001, 0.01, 0.1, and 1 s from top to bottom (the values are listed in the left panel).\label{fig:f2}}
\end{figure}

\begin{figure}
	%\figurenum{4}
	\plotone{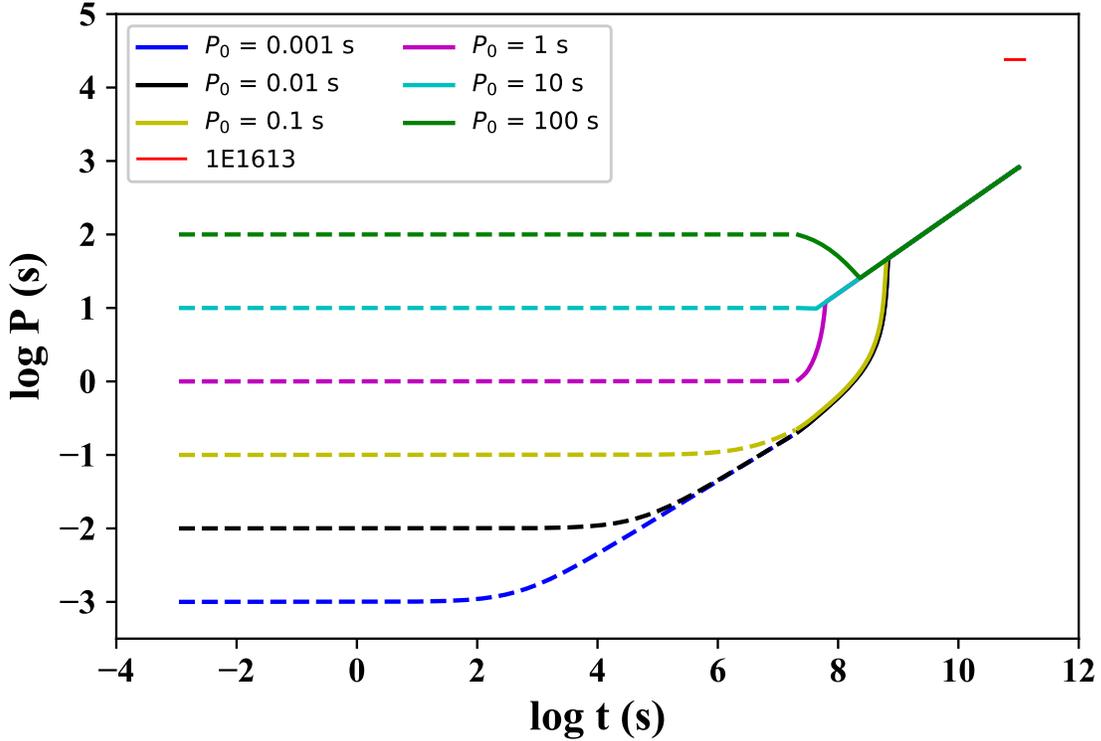}
	\caption{The spin evolution with the initial spin period taken to be 0.001, 0.01, 0.1, 1, 10, and 100 s. In all cases $M_{\rm d} = 10^{-5} M_\odot$ and $R_{\rm f} = 10^4 R_{\rm S}$. The spin periods coincide at the time $\sim 10^9$ s and evolve to $\sim 820$ s at the time 3200 yr. In each case the solid and dashed and solid curves represent the evolutionary stages with and without NS-disk interaction, respectively. The red line represents the period of 1E1613, and its length corresponds to the possible range of the age ($2000-4000$ yr) of RCW 103.\label{fig:f2}}
\end{figure}

\begin{figure}
	\centering
	%\figurenum{5}
	\includegraphics[width=0.9\textwidth]{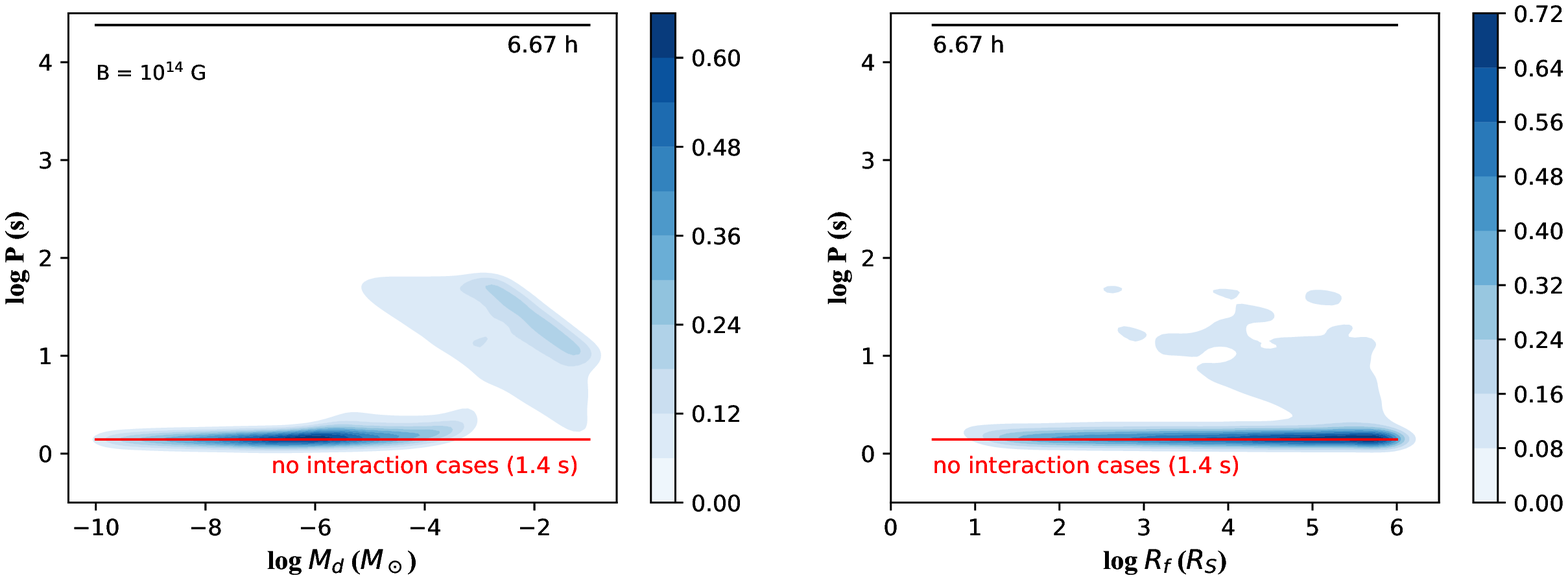}
	\includegraphics[width=0.9\textwidth]{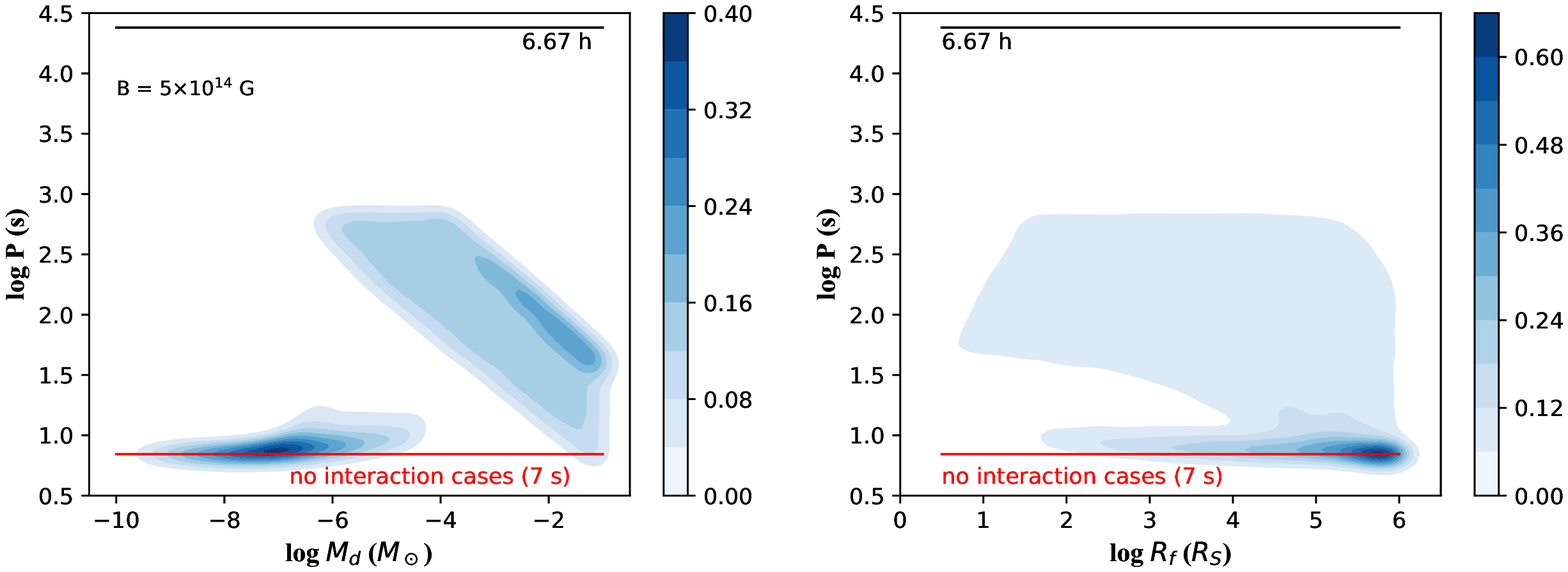}
	\includegraphics[width=0.9\textwidth]{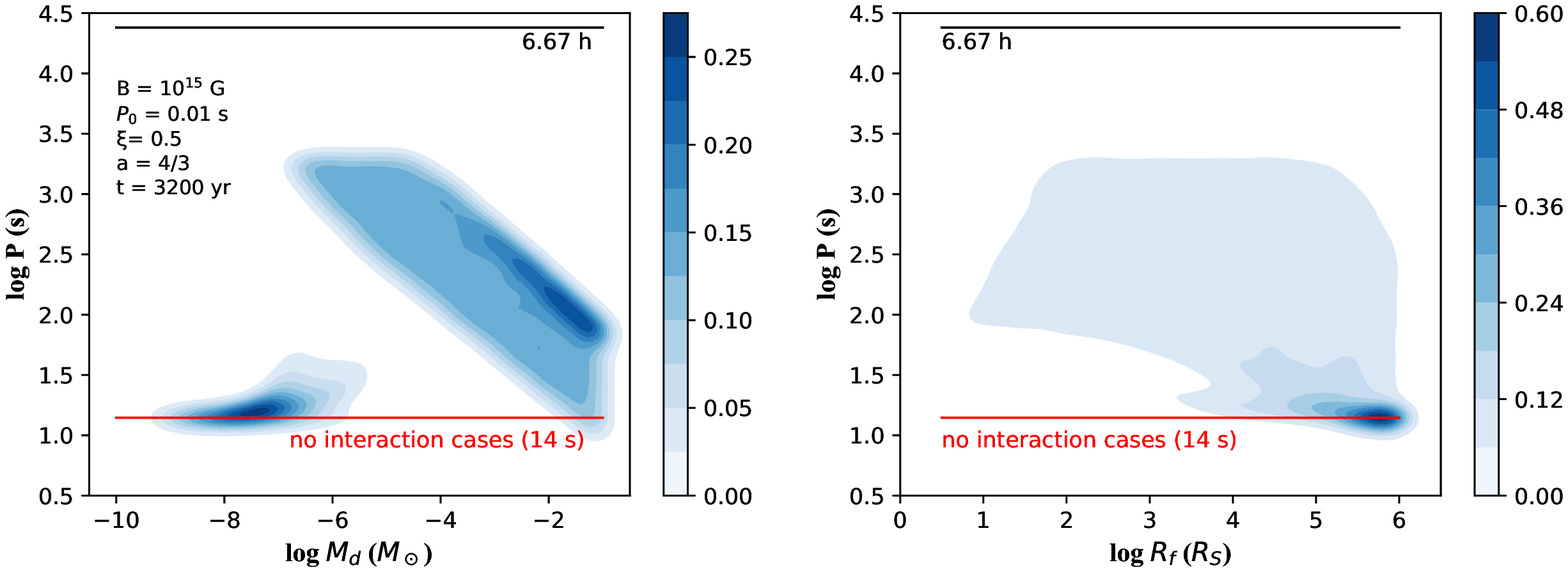}
	\includegraphics[width=0.9\textwidth]{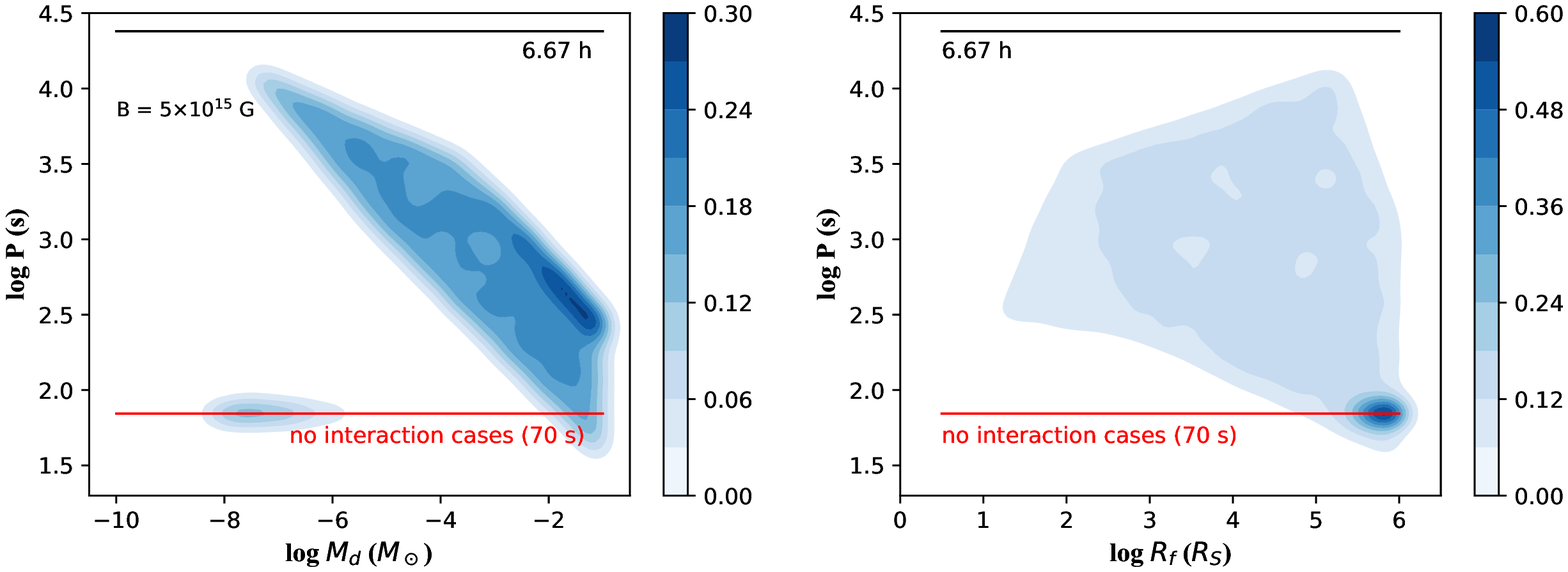}
	\caption{Same as Fig.~2 but with the initial magnetic field $B = 10^{14}$, $5 \times 10^{14}$, $10^{15}$, and $5 \times 10^{15}$ G from top to bottom. The red lines (corresponding to periods of about 1.4 s, 7 s, 14 s and 70 s from top to bottom) represent the cases without disk interaction.}
\end{figure}

\begin{figure}
	%\figurenum{6}
	\plotone{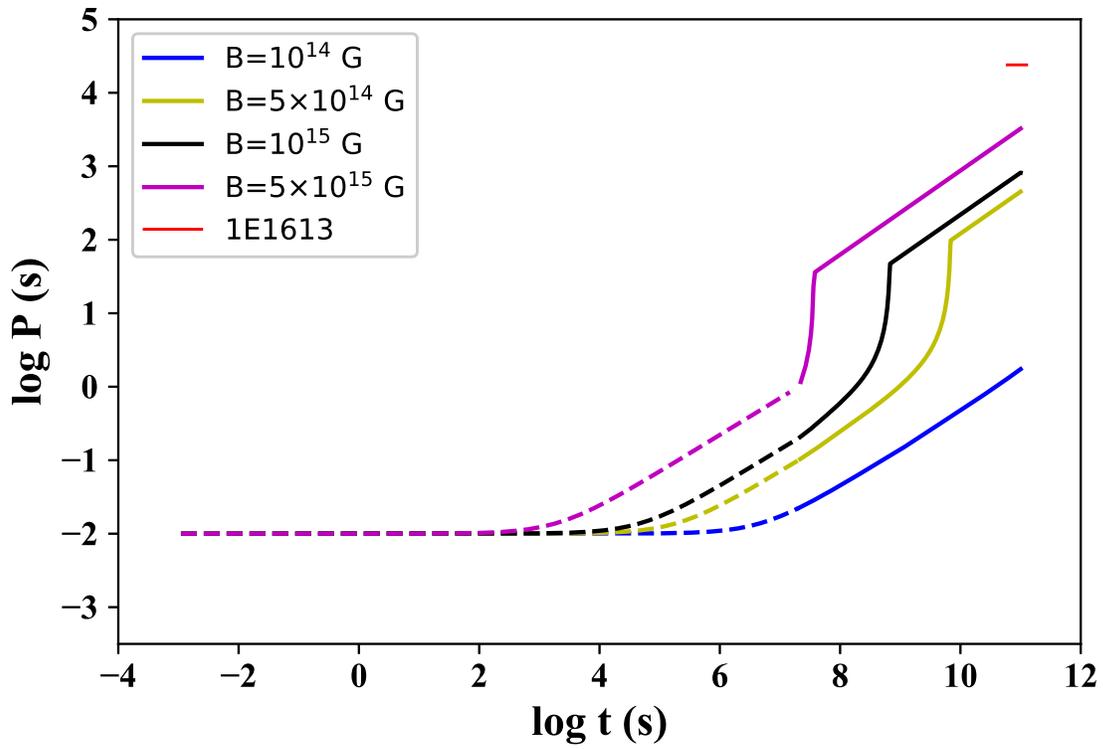}
	\caption{Same as Fig.~4, but for the NS spin evolution with the initial magnetic field $B = 10^{14}$, $5 \times 10^{14}$, $10^{15}$, and $ 5 \times 10^{15}$ G. }
\end{figure}

\begin{figure}
	\centering
	%\figurenum{7}
	\includegraphics[width=1.0\textwidth]{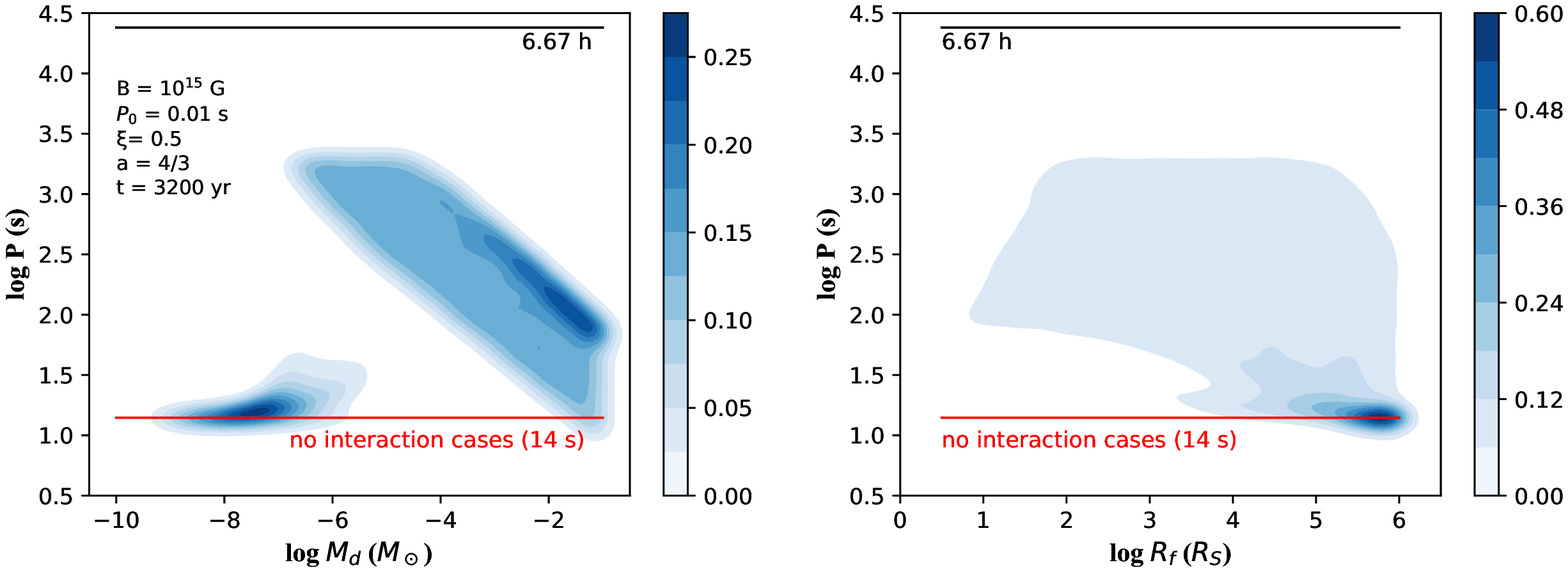}
	\includegraphics[width=1.0\textwidth]{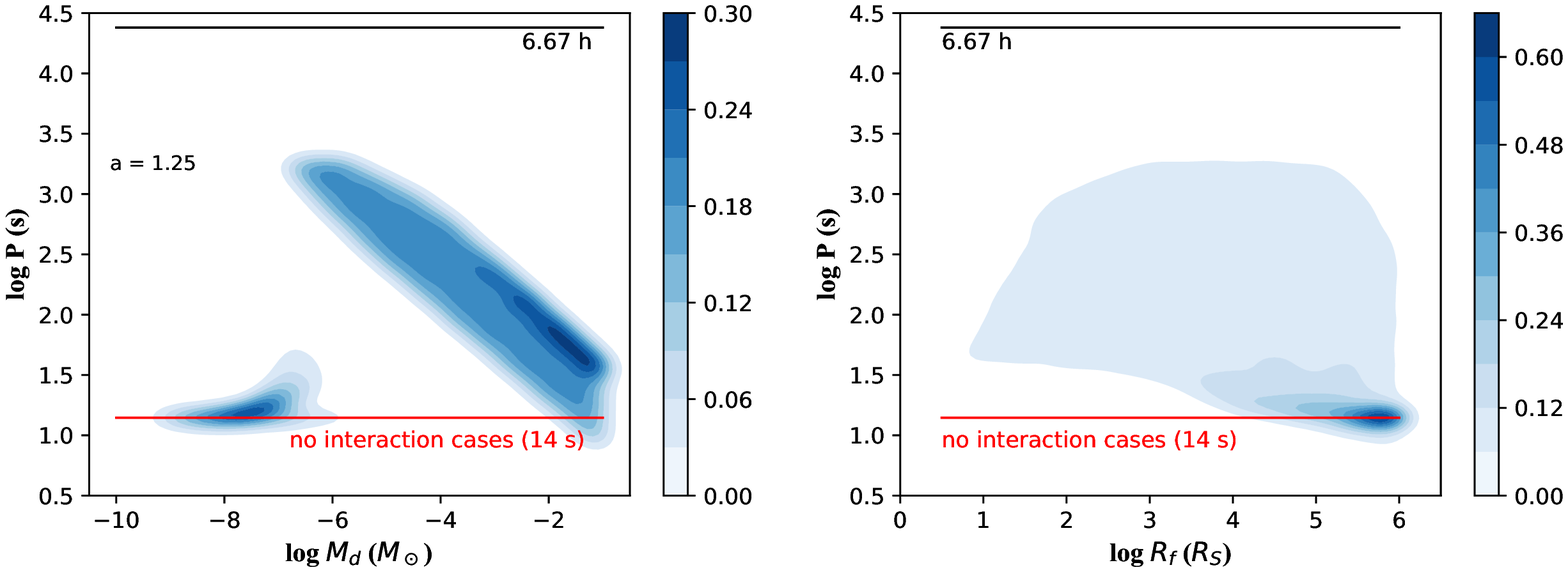}
	\caption{Same as Fig.~2 but with $a = 4/3$ (upper panels) and 1.25 (lower panels).}
\end{figure}

%\begin{figure}
%	\figurenum{8}
%	\plotone{fig9.eps}
%	\caption{A detailed example evolution of two cases with different power index $a$, which are $a = 4/3$ (the black curve) and $a = 1.25$ (the blue curve).\label{fig:f2}}
%\end{figure}

\begin{figure}
	\centering
	%\figurenum{9}
	\includegraphics[width=1.0\textwidth]{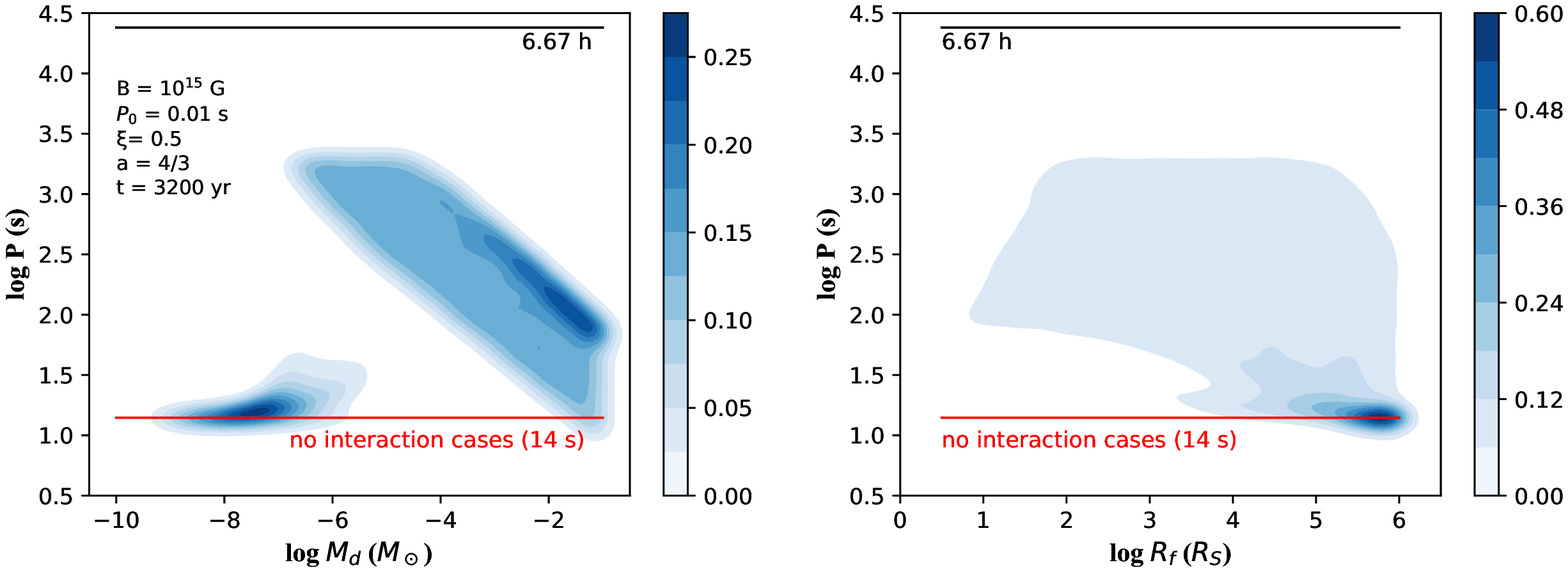}
	\includegraphics[width=1.0\textwidth]{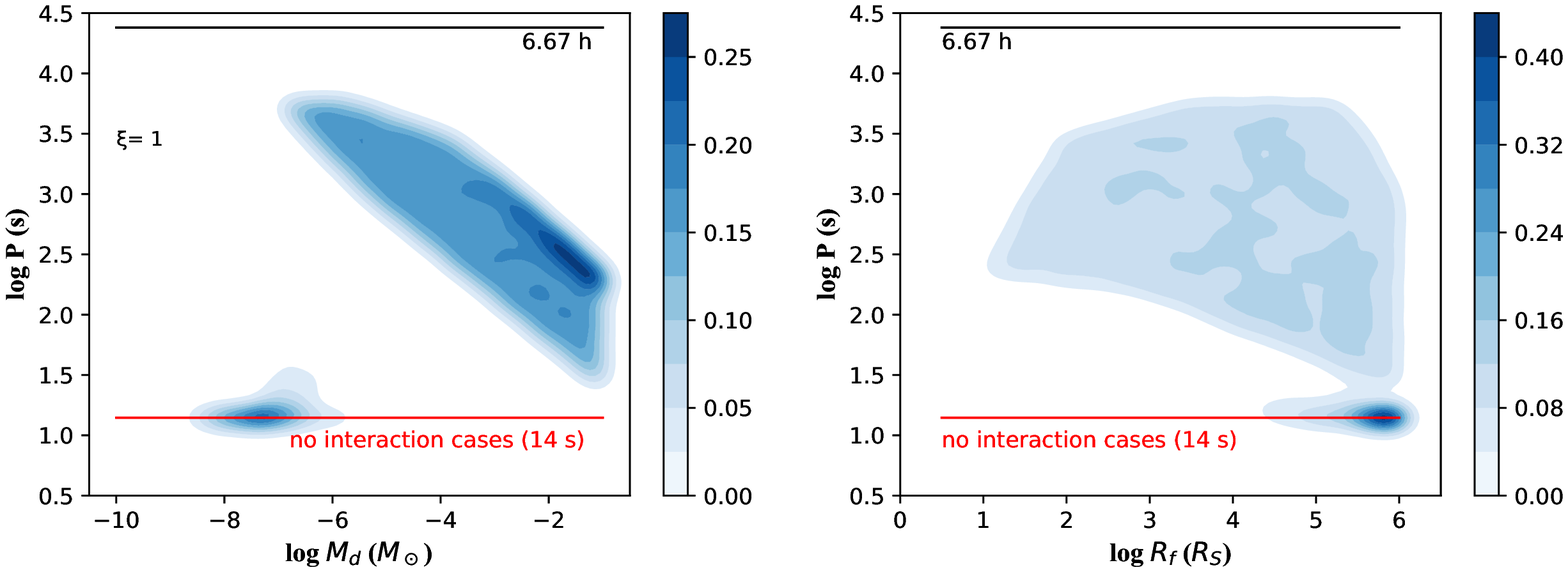}
	\caption{Same as Fig.~2 but with $\xi=0.5$ (upper panels) and 1 (lower panels). }
\end{figure}

%\begin{figure}
%	\figurenum{10}
%	\plotone{fig11.eps}
%	\caption{A detailed example evolution of two cases with different $\xi$, which are $xi = 0.5$ (the black curve) and $\xi = 1$ (the blue curve). \label{fig:f2}}
%\end{figure}

\begin{figure}
	\centering
	%\figurenum{11}
	\includegraphics[width=1.0\textwidth]{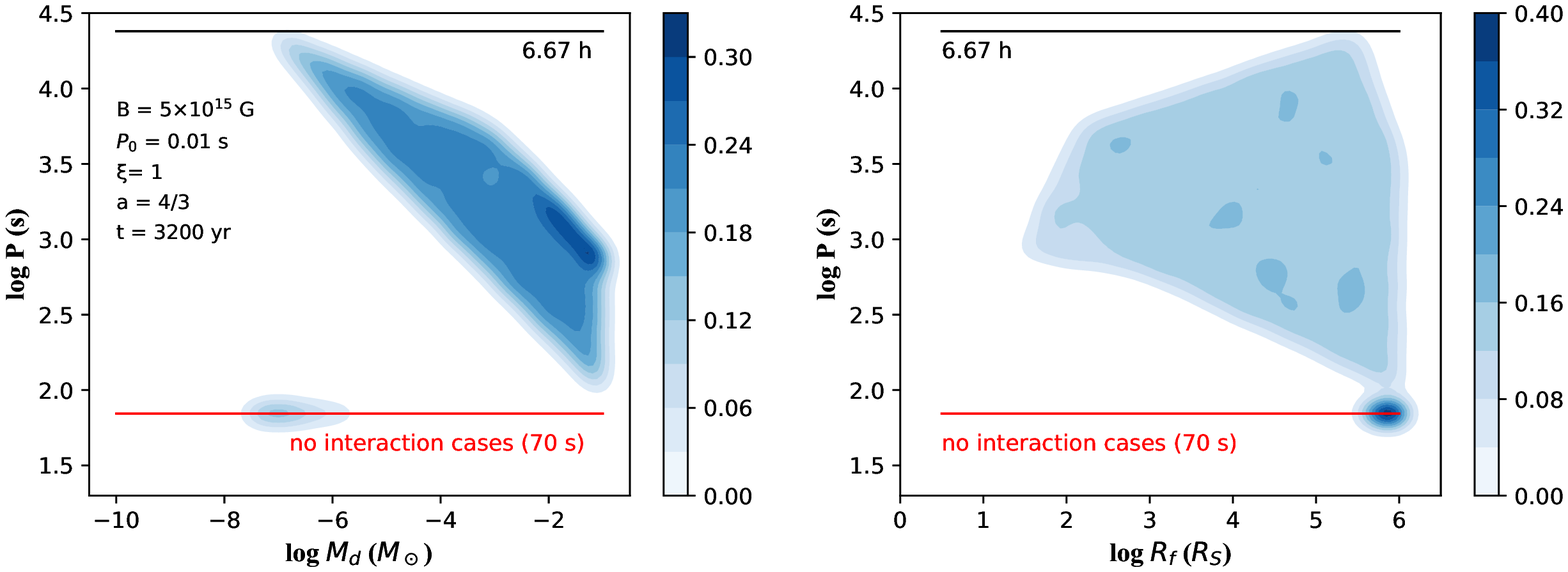}
	\includegraphics[width=1.0\textwidth]{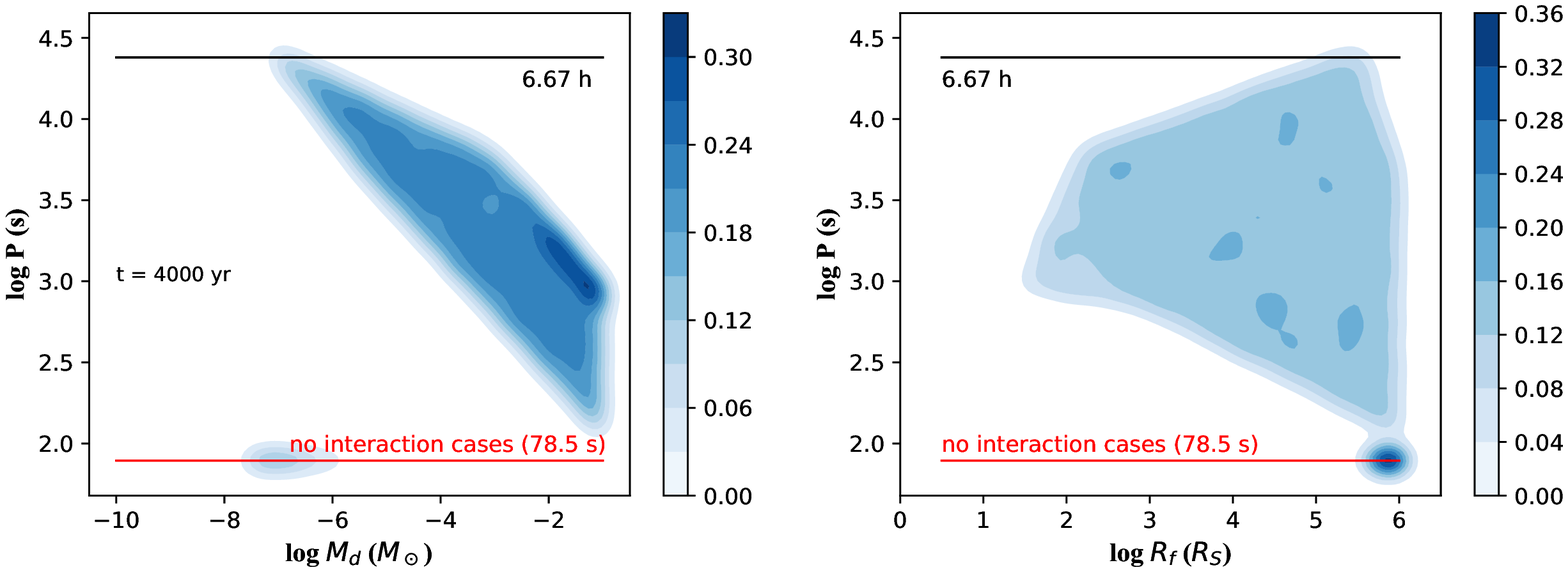}
	\caption{Same as Fig.~2 but with $B=5 \times 10^{15}$ G and $\xi=1$. In the upper and lower panels the ending time is taken to be 3200 and 4000 yr, respectively. The red lines (corresponding to periods of about 70 s and 78.5 s from top to bottom) represent the cases without disk interaction.}
\end{figure}

\begin{figure}
	\centering
	%\figurenum{10}
	\plotone{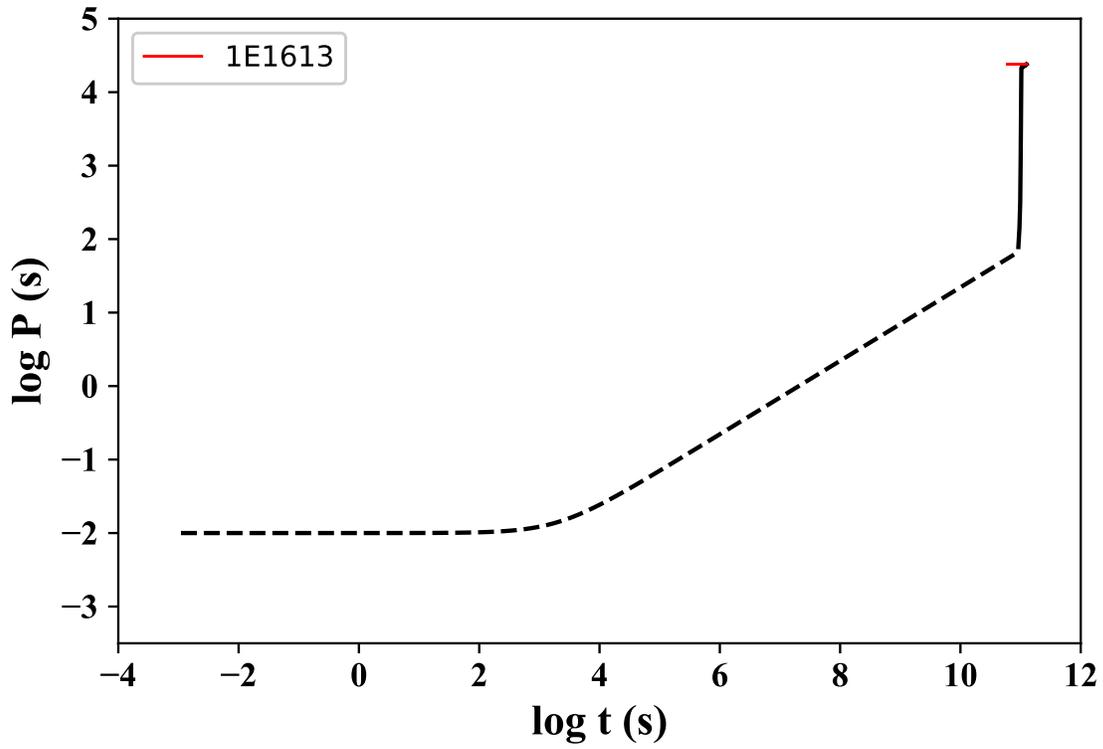}
	\caption{The NS spin evolution with $B=5 \times 10^{15}$ G, $P_0 = 0.01$ s, $\xi=1$, $M_{\rm d}=8.9 \times 10^{-8} M_{\odot}$, and $R_{\rm f} =3.5 \times 10^5 R_{\rm S}$. The final spin period at the age of 4000 yr is 24042 s.}
\end{figure}

\begin{figure}
	\centering
	%\figurenum{11}
	\plotone{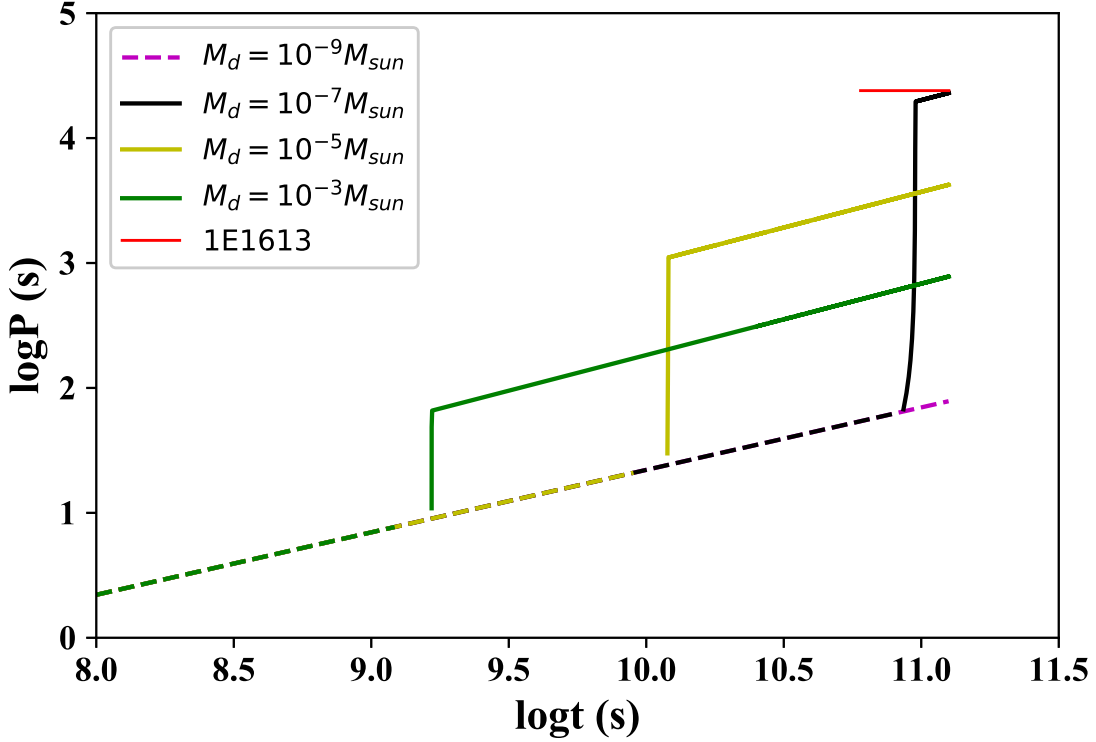}
	\caption{Same as Fig.~10 but with $M_d = 10^{-9} M_{\odot}$, $10^{-7} M_{\odot}$, $10^{-5} M_{\odot}$, and $10^{-3} M_{\odot}$.}
\end{figure}

\begin{figure}
	\centering
	%\figurenum{12}
	%\plotone{fig12.eps}
	\plottwo{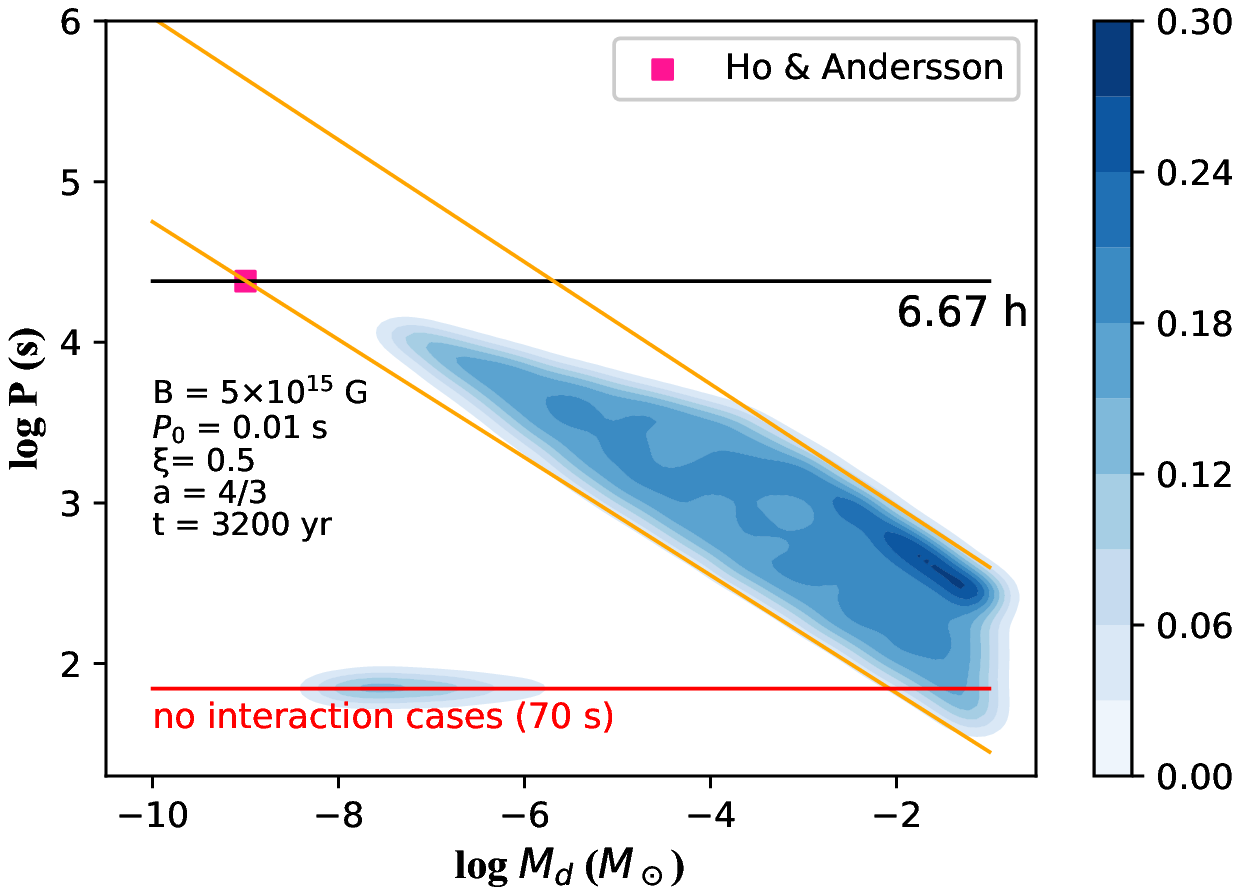}{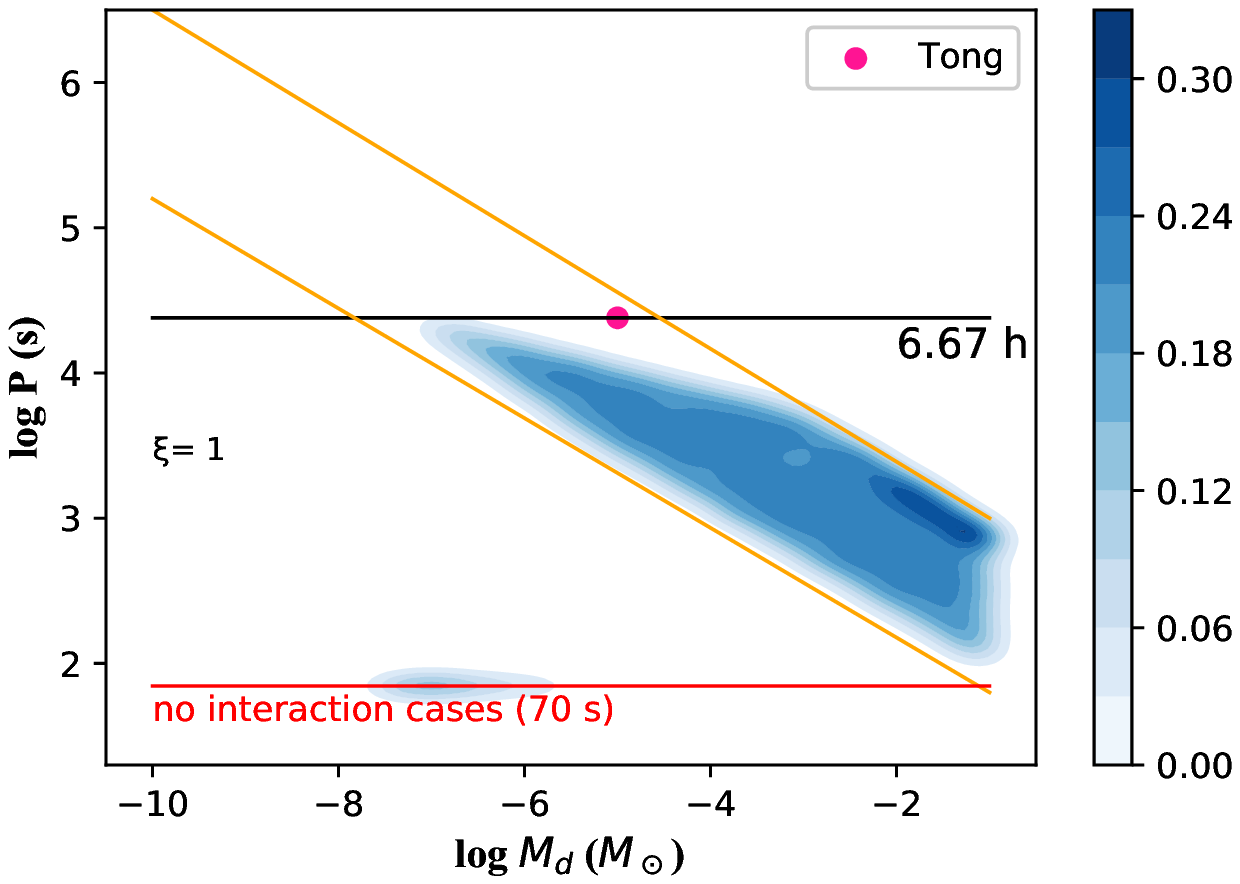}
	\caption{The blue regions represent the distribution of NSs that reach equilibrium spin perids in the $P-M_{\rm d}$ plane.  The red line (with the period around 70 s) represent the cases without disk interaction.
	The two orange lines confining the predicted regions represent the lower and upper boundaries of the equilibrium spin periods.
	The red square in the left panel and the red circle in the right panel correspond to the results of \citet{ha2017} and \citet{twlx2016}, respectively. }
\end{figure}

\clearpage

\label{lastpage}
\end{document}